\documentclass{jfm}
\usepackage{amssymb}
\usepackage{natbib}
\usepackage{lineno}
\usepackage{amsfonts,amsbsy}
\usepackage{graphicx}
\usepackage{color}

\linenumbers

\def\e{\mathrm{e}}

\def\d{\mathrm{d}}

\def\beq{\begin{equation}}
\def\eeq{\end{equation}}

\def\bx{\boldsymbol{x}}

\def\bu{\boldsymbol{u}}

\def\bvarphi{\boldsymbol{\varphi}}
\def\bxi{\boldsymbol{\xi}}

\def\bq{\boldsymbol{q}}
\def\bX{\boldsymbol{X}}
\def\Pe{\mathrm{Pe}}
\def\E{\mathbb{E}\,}
\def\tbq{\tilde{\bq}}
\def\f{\mathfrak{f}}
\def\f{{\mathsf{f}}}
\def\keff{\mathsf{k}}

\newcommand{\dpar}[2]{\frac{\partial #1}{\partial #2}}

\newcommand{\dt}[2]{\frac{\mathrm{d} #1}{\mathrm{d} #2}}

\newcommand{\eqn}[1]{(\ref{eqn:#1})}
\newcommand{\lab}[1]{\label{eqn:#1}}
\newcommand{\inter}[1]{\quad \textrm{#1} \quad}

\def\XXint#1#2#3{{\setbox0=\hbox{$#1{#2#3}{\int}$}
\vcenter{\hbox{$#2#3$}}\kern-.5\wd0}}

\title[Dispersion in the large-deviation regime II]{Dispersion in the large-deviation regime. Part II: cellular flow at large P\'eclet number}
\author[P. H. Haynes and J. Vanneste]{P. H. Haynes$^1$ and J. Vanneste$^2$\thanks{Email address for correspondence: J.Vanneste@ed.ac.uk}}
\affiliation{$^1$Department of Applied Mathematics and Theoretical Physics, University of Cambridge, Wilberforce Road, Cambridge CB3 0WA, UK \\[\affilskip] 
$^2$School of Mathematics and Maxwell Institute for Mathematical Sciences, University of Edinburgh, King's Buildings, Edinburgh EH9 3JZ, UK}

\begin{document}

\maketitle

\begin{abstract}
A standard model for the study of scalar dispersion through the combined effect of advection and molecular diffusion is a two-dimensional periodic flow with closed streamlines inside periodic cells. Over long time scales, the dispersion of a scalar released in this flow can be characterised by an effective diffusivity that is a factor $\Pe^{1/2}$ larger than molecular diffusivity  when the P\'eclet number $\Pe$ is large. Here we provide a more complete description of dispersion in this regime by applying the large-deviation theory developed in Part I of this paper. Specifically, we derive approximations to the rate function governing the scalar concentration at large time $t$ by carrying out an asymptotic analysis of the relevant family of eigenvalue problems. 

We identify two asymptotic regimes and, for each, make predictions for the rate function and spatial structure of the scalar. Regime I applies to distances $|\bx|$ from the scalar release point that satisfy $|\bx| = O(\Pe^{1/4} t)$ . The concentration in this regime is isotropic at large scales, is uniform along streamlines within each cell, and varies rapidly in boundary layers surrounding the separatrices between adjacent cells. The results of homogenisation theory, yielding the $O(\Pe^{1/2})$ effective diffusivity, are recovered from our analysis in the limit $|\bx| \ll \Pe^{1/4} t$. Regime II applies when $|\bx|=O(\Pe \, t/\log \Pe)$ and is characterised by an anisotropic concentration distribution that is localised around the separatrices. A novel feature of this regime is the crucial role played by the dynamics near the hyperbolic stagnation points.
A consequence is that in part of the regime the dispersion can be interpreted as  resulting from a random walk on the lattice of stagnation points.
The two regimes overlap so that our asymptotic results describe the scalar concentration over a large range of distances $|\bx|$. They are verified against numerical solutions of the family of eigenvalue problems yielding the rate function.

\end{abstract}

\section{Introduction}

\begin{figure}
\begin{center}
\includegraphics[height=5cm]{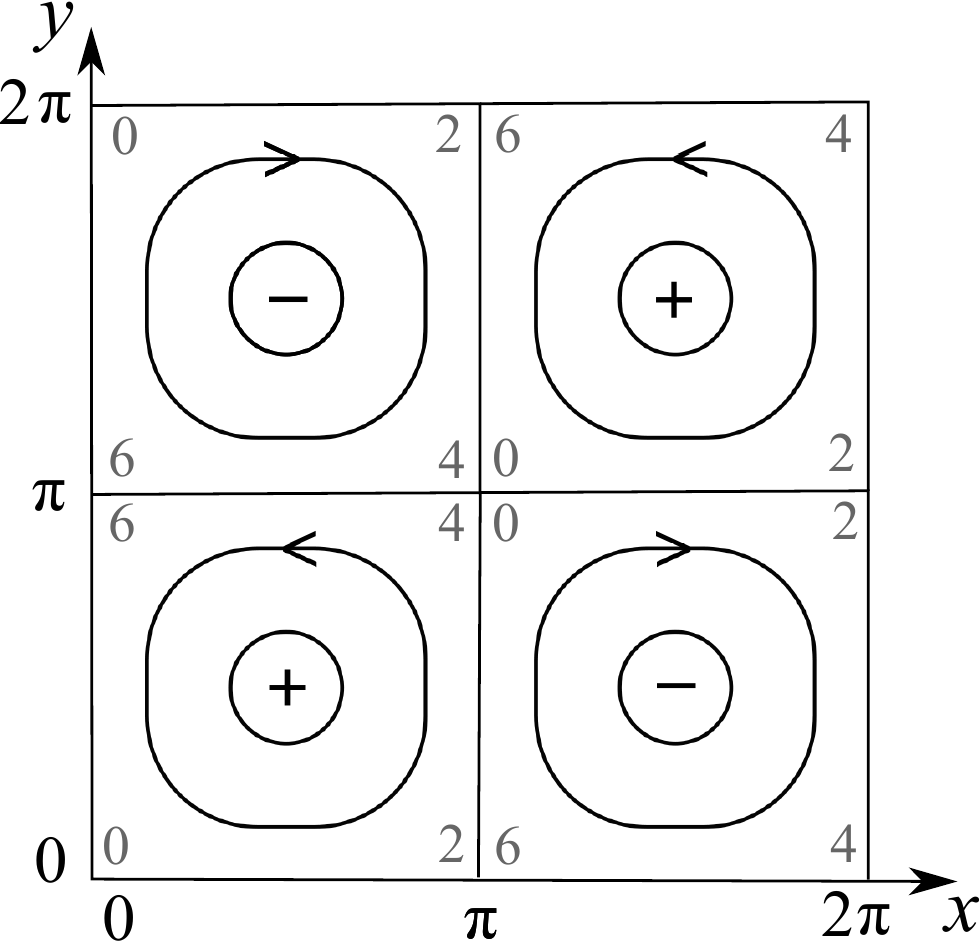} 
\caption{Schematic of the streamlines in single periodic cell for the flow with streamfunction \eqn{psinodim}. The quarter-cells with anticlockwise (clockwise) circulation are denoted by the $+$ ($-$) signs. The value of the coordinate $\sigma$ used in the boundary-layer analyses of \S\S\ref{sec:Ibl} and \ref{sec:II} is indicated at the corner of each quarter-cell by the grey number.}
\label{fig:cellpicture}
\end{center}
\end{figure}

The transport and mixing of constituents by fluid flows is a classical problem in fluid dynamics, motivated by a broad range of industrial and environmental applications. One of the main strands of the research on this problem, e.g.\ reviewed in \citet{majd-kram}, examines how the interaction between advection and molecular diffusion leads to enhanced transport and mixing.  
The dispersion resulting from advection and molecular diffusion gives, in the long-time limit, both linearly increasing variance of particle positions and Gaussian concentration distributions and can therefore be quantified by means of an  effective diffusivity (typically much increased compared to the molecular value). Several approaches---including homogenisation---are available to calculate this effective diffusivity and, in simple examples at least, they yield instructive closed-form results \citep{majd-kram}. Classical examples of this type are shear flows, originally considered by \citet{tayl53}, and the cellular flow on which the present paper focuses. The cellular flow is a two-dimensional periodic incompressible flow, with streamfunction
\beq \lab{psi}
\psi = - U a \sin(x/a) \sin (y/a),
\eeq
where $U$ is the maximum flow speed and $2\pi a$ is the cell period. This flow consists of a doubly infinite array of periodic cells in which the fluid is rotating alternatively clockwise and anti-clockwise as sketched in Fig.~\ref{fig:cellpicture}. It was introduced in studies of kinematic dynamos \citep{chil79} and has since become a benchmark for work on advection--diffusion \citep[e.g.][]{moff83}. The enhancement of dispersion by this flow (over that which results from molecular diffusion alone) is encapsulated by the results of \citet{sowa87}, \citet{shra87} and \citet{rose-et-al} showing that the  effective diffusivity in this case scales like $\Pe^{1/2}$ as $\Pe \to \infty$. Here 
\beq \lab{peclet}
\Pe = U a/\kappa, 
\eeq
with $\kappa$ the molecular diffusivity, is the P\'eclet number, which measures the relative strength of advection and diffusion in the flow. Several other explicit results are also available for this flow, see \citet{majd-kram} and references therein.

When applied to initial-value problems, which typically involve a passive scalar released  initially in a small region, the characterisation of dispersion by a single effective diffusivity relies on an implicit assumption: the distances $| \bx|$ between  points of interest and the scalar-release region are assumed to be moderately large, specifically to be $O(t^{1/2})$ for large times $t$. For larger distances, the approximation by a diffusive, Gaussian process is invalid. In a companion paper \citep[][hereafter referred to as Part I]{hayn-v14a} we show that the scalar distribution at such distances can nonetheless by described analytically, using the theory of large deviations. Part I provides a general formulation for the theory of large deviations relevant to dispersion problems and applies it to shear flows and to periodic flows including the cellular flow \eqn{psi}. The results presented there for the cellular flow are largely numerical although asymptotic expressions are obtained in the regime $\Pe \ll 1$ corresponding to weak advection. In the present paper we obtain detailed asymptotic results for the opposite, and arguably more physically interesting, regime $\Pe \gg 1$ corresponding to weak diffusion. In this limit, diffusion acts as a small, singular perturbation to advection: in the complete absence of diffusion, there is no large-scale transport since particle trajectories are confined to closed streamlines inside quarter cells (see Figure \ref{fig:cellpicture}). A weak diffusion makes it possible for particles to migrate from streamline to streamline and, when crossing the separatrices, from cell to cell, leading to large-scale transport. The importance of the separatrices is reflected in the asymptotic analysis which largely consists of a boundary-layer treatment of an $O(\Pe^{-1/2})$ region surrounding them.

Specifically, we identify and study two asymptotic regimes, characterised by different scalings of $|\bx|/t$ relative to $\Pe$ and leading to different asymptotic reductions. The boundary-layer analysis in the first regime turns out to be same as that appearing in the computation of the $O(\Pe^{1/2})$ effective diffusivity \citep{chil79,sowa87}, even though the regime
applies over a broader range of $|\bx|/t$ and captures non-diffusive effects. The boundary-layer analysis in the second case is novel. It requires a careful treatment of the regions around the stagnation points, leading to a logarithmic dependence of the results on $\Pe$. Note that in both cases the asymptotic analysis is formal: we make no attempt at bounding error terms. Instead, we  check our asymptotic predictions against numerical solutions of the eigenvalue problem determining the scalar concentration in the large-deviation regime, and we find excellent agreement.

\begin{figure}
\begin{center}
\includegraphics[height=5cm]{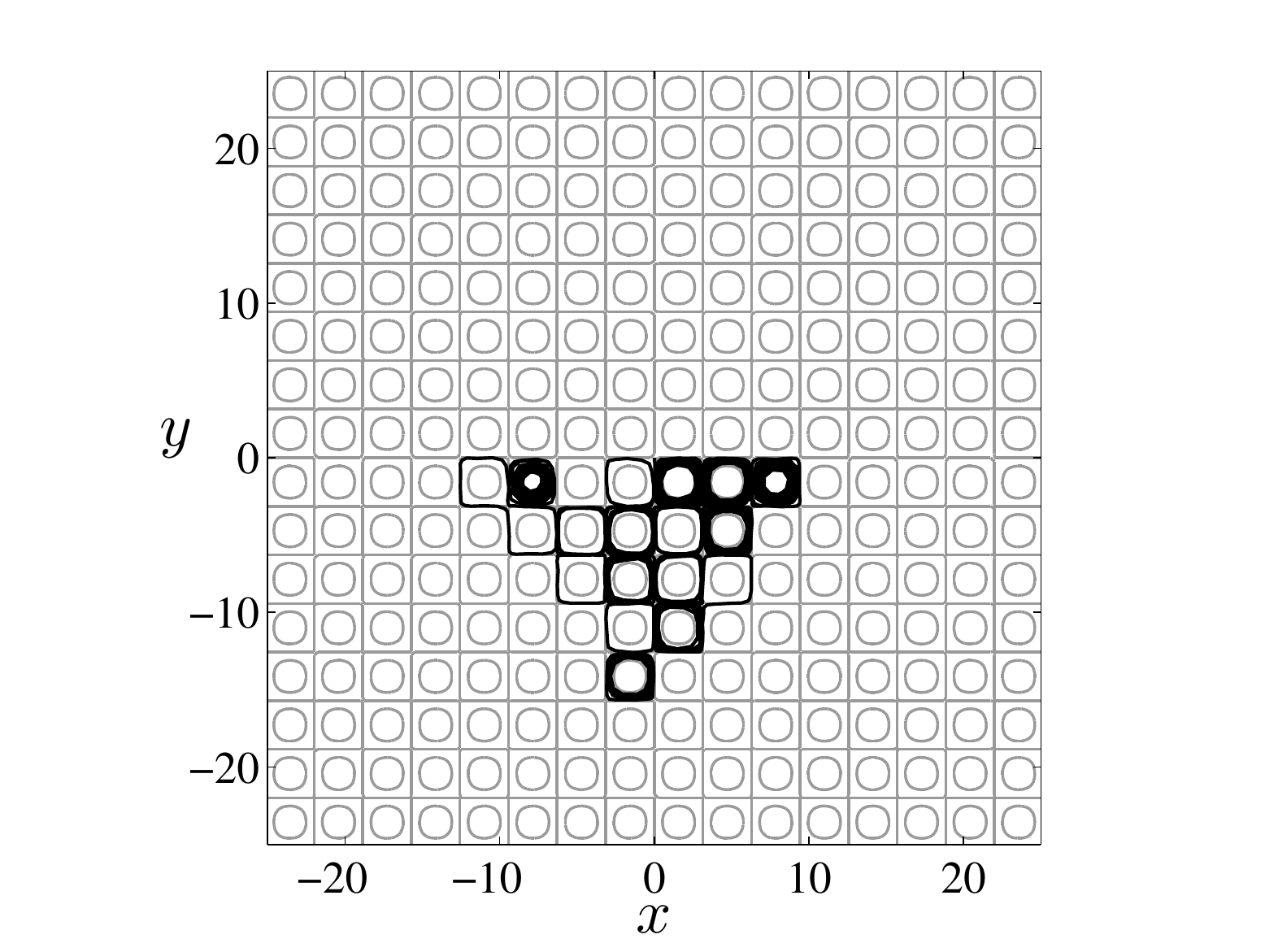}  \hspace{-1cm} \includegraphics[height=5cm]{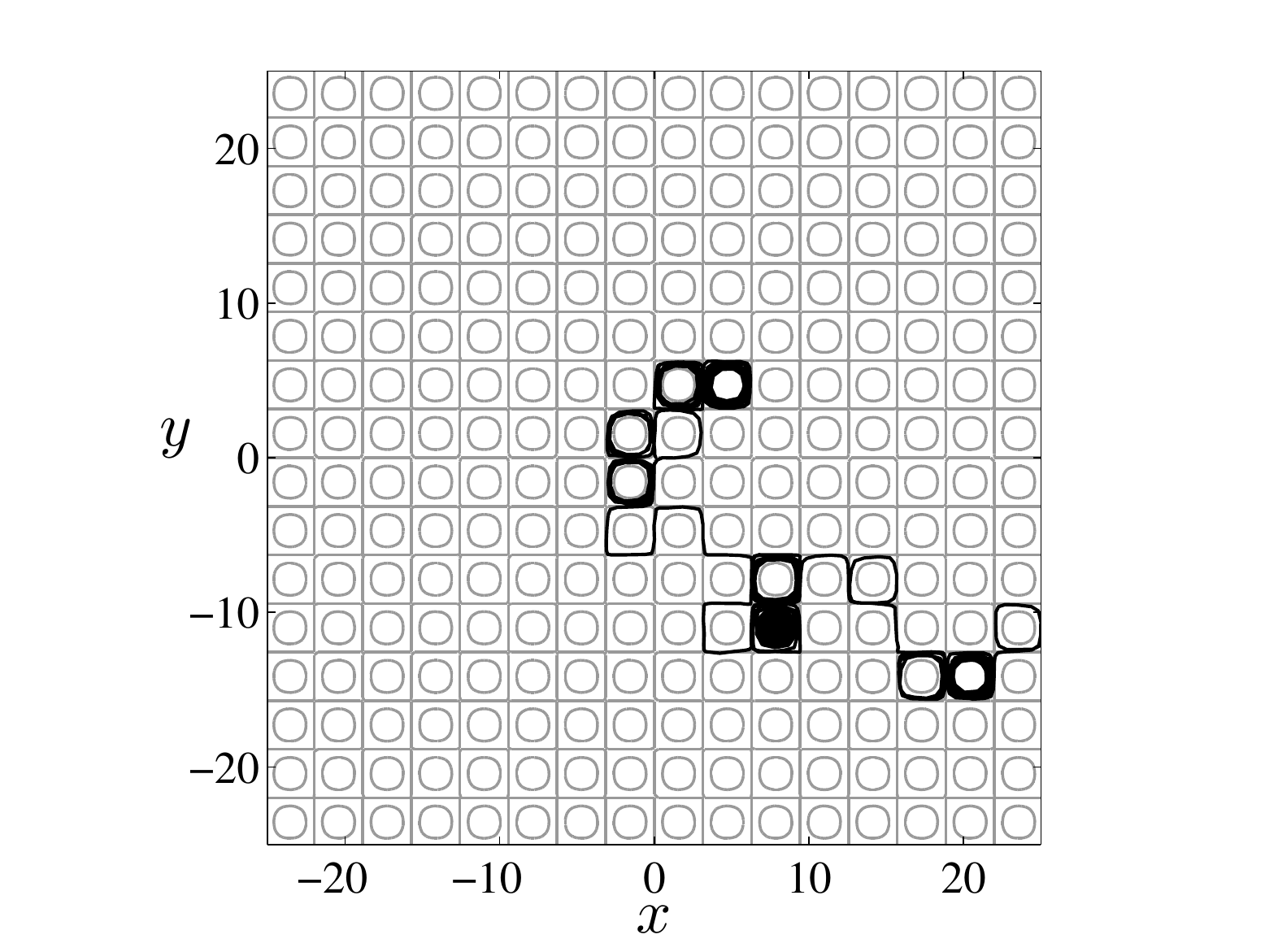}
\caption{Examples of particle trajectories for $\Pe=1000$. The black lines show the trajectories $\bx(t)$ for $0 \le t \le 2000$ of two particles initially released at the origin for two different realisations of the Brownian motion. The grey lines indicate the streamlines $\psi=0$ (separatrices) and $\psi=\pm 1/2$.}
\label{fig:trajplot}
\end{center}
\end{figure}

Before entering the intricacies of this asymptotic analysis, however, it is useful to have in mind a physical picture of dispersion in the cellular flow at high $\Pe$. Figure 5 of Part I illustrates the dispersion by displaying the evolution of the concentration of a passive scalar released in the central cell. It indicates a transition, moving away from the release region,  between cells that have near-uniform concentrations near the centre and cells that are depleted at larger distances, with non-zero concentration essentially confined to the neighbourhood of the separatrices. The asymptotic analysis presented in this paper captures this transition  and provides the analytic form of the concentration distribution over a broad range of distances. An alternative view of the problem considers independent particles released in the flow \eqn{psi} and experiencing different realisations of the Brownian motion associated with diffusion. In this view, the (normalised) scalar concentration is interpreted as the particle-position probability distribution function. The motion of single particles is illustrated in Figure \ref{fig:trajplot} showing two trajectory realisations. As expected, most of the time particles are trapped within quarter cells for long times; as the right panel suggests, however, large excursions are possible when the Brownian motion is such that the particle remains close to the separatrix for some time. The statistics at large distances from the release point are then controlled by rare realisations of the Brownian motion for which the particle only rarely visits the cell interiors. The large-deviation theory we employ is the probabilistic tool required to capture the statistics of these rare realisations.

\section{Formulation} \label{sec:formulation}

We examine the dispersion of a passive scalar in the cellular flow with streamfunction \eqn{psi}. The concentration $C(\bx,t)$ of a passive scalar released in this flow is governed by the advection--diffusion equation. Using $a$ as reference length and the  diffusive time scale $a^2/\kappa$ as reference time, this equation takes the non-dimensional form
\beq \lab{ad-dif}
\partial_t C + \Pe \, \bu \cdot \nabla C = \nabla^2 C,
\eeq
where $\bu=(-\partial_y \psi,\partial_x \psi)$ and 
\beq \lab{psinodim}
\psi=-\sin x \sin y
\eeq
are the dimensionless velocity and streamfunction. We consider the initial-value problem with initial condition $C(\bx,0)=\delta(\bx)$ so that $C(\bx,t)$ can interpreted as a particle-position probability density function for a particle released at the origin. 

In Part I, we show that the concentration for $t \gg 1$ takes the large-deviation form
\beq \lab{largedevi}
C(\bx,t) \sim t^{-1} \phi(\bx,\bxi) \e^{-t g(\bxi)},\quad \textrm{where} \ \ \bxi = \bx/t.
\eeq
The Cram\'er or rate function $g(\bxi)$ which appears in \eqn{largedevi} controls the dispersion for $\bx = O(t)$ and is our main object of interest. It can be determined as the Legendre transform of the dual function $f(\bq)$,
%defined by 
%\beq \lab{f}
%f(\bq) = \lim_{t \to \infty} \frac{1}{t} \log 
%\eeq
identified as $t^{-1}$ times the cumulant generating function $\log \E \e^{\bq \cdot \bX}$ for the position $\bX$ of particles advected and diffused in the flow, where $\E$ denotes expectation over the Brownian motion associated with diffusion. The relationship between $g(\bxi)$ and $f(\bq)$ follows from the Ellis--G\"artner theorem, a key result of large-deviation theory \citep[e.g.][]{elli95,demb-zeit,denh00,touc09}. In turn, $f(\bq)$ is found as the principal eigenvalue of the eigenvalue problem
\beq \lab{eig1}
 \nabla^2 \phi - \left(\Pe \, \bu + 2 \bq \right) \cdot \nabla \phi + \left(\Pe \,\bu \cdot \bq + |\bq|^2\right) \phi = f(\bq) \phi,
\eeq
where $\bq=(q_1,q_2)$ is regarded as a parameter and $\phi(\bx,\bq)$ is the eigenfunction which satisfies periodic boundary conditions. 

The principal eigenvalue of \eqn{eig1} is guaranteed to be real, with a corresponding eigenfunction $\phi$ that is real and sign definite (see Part I). This eigenfunction describes the spatial structure of the concentration: according to \eqn{largedevi} and the relation $\bq=\nabla_{\bxi} g$, the structure of $C(\bx,t)$ at fixed $t$ is locally proportional to $\phi(\bx,\bx t) \exp(-\bq \cdot \bx)$. Note that the eigenvalue problem \eqn{eig1} also appears  when the Floquet--Bloch theory of differential equations with periodic coefficients is applied to \eqn{ad-dif}  (see \citealt{bens-et-al}, \S4.3.1;  \citealt[][\S3.6]{papa95}), and in the problem of front propagation in the presence of an FKPP chemical reaction \citep[see][]{novi-ryzh,xin09}.

In Part I, we examine  scalar dispersion in cellular flows by solving \eqn{eig1} numerically for fixed $\Pe$ for a range of values $\bq$ and then deducing $g(\bxi)$  by Legendre transform. 
We also provide an asymptotic approximation to $f(\bq)$ in the limit $\Pe \to 0$. Here we obtain a detailed description in the opposite limit $\Pe \to \infty$. This limit has received a great deal of attention, most of which has been devoted to the determination of an effective diffusivity $\keff$  \citep{chil79,shra87,rose-et-al,sowa87,fann-papa,kora04,novi-et-al,gorb-et-al}. This characterises the dispersion for $\bx=O(t^{1/2})$ by providing the (Gaussian) diffusive approximation $C(\bx,t) \asymp \exp\left(-|\bx|^2/(4 \keff t) \right)$ to the concentration. The key result in this area is the asymptotic approximation
\beq \lab{effdiff}
\keff \sim 2 \nu \Pe^{1/2}
\eeq
\citep{chil79,shra87,rose-et-al,sowa87},
where the constant 
\beq \lab{nu}
\nu = \left(\frac{2}{\pi}\right)^{1/2} \sum_{n=0}^\infty \frac{(-1)^n}{(2n+1)^{1/2}} = 0.532740705\cdots
\eeq
was determined by \citet{sowa87}. This result is obtained by applying a boundary-layer analysis to the so-called cell problem which arises when computing an effective diffusivity using the method of homogenisation \citep[e.g.][]{majd-kram}. 
Since the effective diffusivity can be deduced from the large-deviation functions $f$ and $g$, specifically from their Taylor expansions
\beq \lab{quadapp}
f(\bq) \sim \keff |\bq|^2 \inter{and} g(\bxi) \sim |\bxi|^2/(4 \keff)
\eeq
for small $\bq$ or $\bxi$ (see Part I), this result is recovered in our large-deviation treatment. 

\begin{figure}
\begin{center}
\includegraphics[height=5cm]{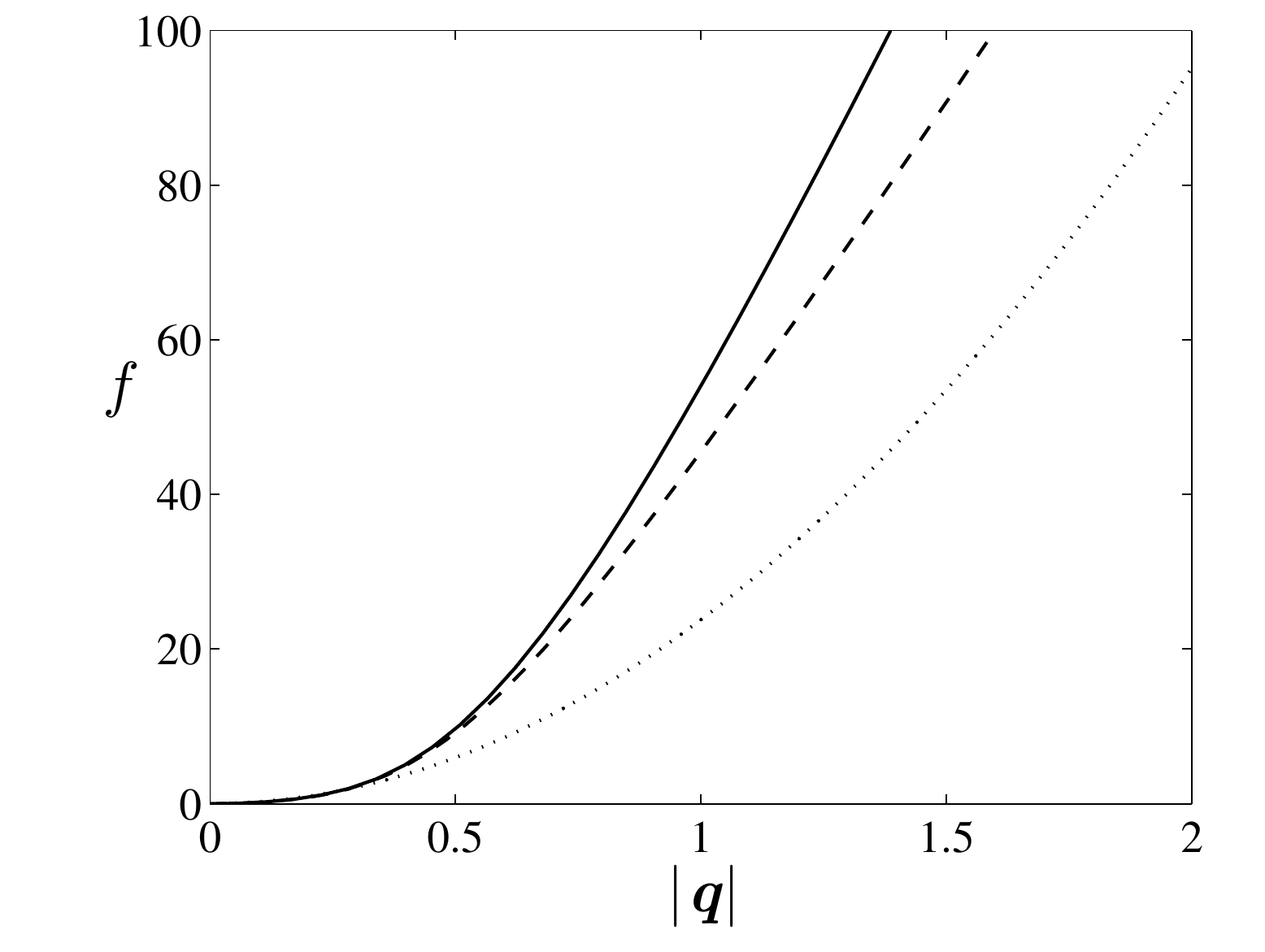}  \includegraphics[height=5cm]{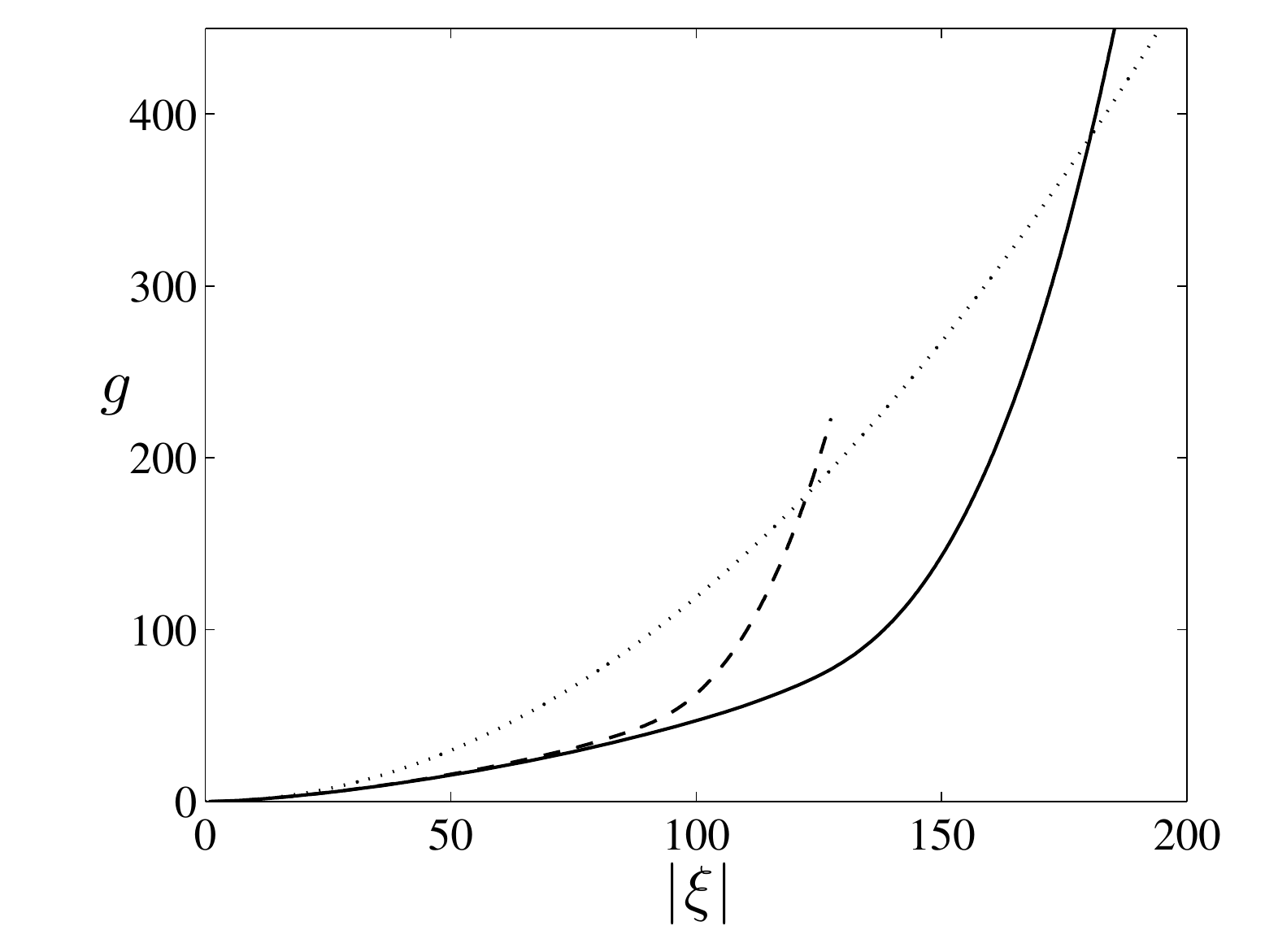} 
\caption{Eigenvalue $f(\bq)$ and rate function $g(\bxi)$ for the cellular flow  with $\Pe=250$ compared with the diffusive approximation \eqn{quadapp}. Left: $f(\bq)$ as a function of $|\bq|$ for $\bq=|\bq|(1,1)/\sqrt{2}$ (solid line) and $\bq=|\bq|(1,0)$ (dashed line). Right: $g(\bxi)$ as a function of $|\bxi|$ for $\bxi=|\bxi|(1,1)/\sqrt{2}$ (solid line) and $\bxi=|\bxi|(1,0)$ (dashed line). The diffusive approximation is shown by the dotted lines.}
\label{fig:fg250}
\end{center}
\end{figure}

To illustrate the limitations of the diffusive approximation, we show in Figure \ref{fig:fg250} the functions $f(\bq)$ and $g(\bxi)$ computed by numerically along  straight lines in each of the $\bq$ and $\bxi$ planes for $\Pe=250$ (these curves correspond to cross sections of Figure 10 in Part I). Both $f(\bq)$ and $g(\bxi)$ differ strikingly from the parabolas of the diffusive approximation with: a clear anisotropy indicating faster dispersion along the diagonal in the $\bx$ plane than along the axes; a $g$ that is broader than parabolic in a intermediate range of $\bx$, corresponding to concentrations exponentially larger than those predicted by diffusion; and a sharp increase in $g$ for large $\bxi$, corresponding to a localisation of the concentration. The asymptotic theory presented in this paper explains these features.

Our derivation of the asymptotic form of $f(\bq)$ for $\Pe \gg 1$ relies on a boundary-layer analysis of \eqn{eig1}. Compared with that leading to the effective diffusivity  \eqn{effdiff}, this derivation is complicated by the presence of the parameter $\bq$. We identify two different regimes, which we denote as I and II characterised by $|\bq|=O(\Pe^{-1/4})$ and $|\bq|=O(1)$, respectively. Regime I is suggested by the approximation \eqn{effdiff}, which implies that $f(\bq) \propto |\bq|^2 = O(\Pe^{-1/2})$ for $|\bq| \ll 1$. In this regime, the eigenvalue $f(\bq)$ is $O(1)$ and is determined by  matching a non-trivial solution in the interior of the flow cells with a boundary-layer solution along the separatrices dividing the cells. The boundary-layer problem for the whole of Regime I turns out to be identical to that arising in the homogenisation approach and solved by \citet{sowa87}. However, it is only in the limit $\Pe^{-1/4} |\bq| \to 0$ that the homogenisation solution, with $\phi$ constant in the cell interiors, and the results \eqn{effdiff}--\eqn{quadapp} are recovered.  
In regime II, $\phi$ vanishes to leading order in the cell interior, and the eigenvalue problem is entirely controlled by the behaviour in the boundary layers. It turns out that $f(\bq)=O(\Pe/\log \Pe)$ in this case. A third regime, expected to arise for $|\bq|=O(\Pe)$, is not considered here since it corresponds to exceedingly small concentrations. It is commented upon the the conclusion of the paper.

We derive the solution $f(\bq)$ in regimes I and II in sections \S\S\,\ref{sec:I}--\ref{sec:II}.
The implications for the rate function $g(\bxi)$ are presented in \S\,\ref{sec:rate}; this provides a more direct physical interpretation of the results since $g(\bxi) \sim t^{-1} \log C(\bx,t)$ (see Part I, \S2.2, for some remarks on the qualitative links between $f(\bq)$ and $g(\bxi)$).
The  paper concludes with a brief discussion in \S\ref{sec:conc}. Throughout the paper we use the following notational convention.
In the period $[0,2\pi]\times [0,2\pi]$ of the flow, two types of quarter cells of size $\pi \times \pi$ need to be distinguished, depending on whether the flow circulates in the positive or negative direction (see Figure \ref{fig:cellpicture}). We denote by $+$ the first type, and by $-$ the second; when using $\pm$ or $\mp$ for expressions with opposite signs in the different cells, the upper (lower) sign refers to + ($-$) cells (so that $\psi=\mp1$ at the centre of cells).

\section{Regime I: $|\bq|=O(\Pe^{-1/4})$} \label{sec:I}

To analyse this first regime, we introduce $\tbq = \Pe^{1/4} \bq$ assumed to be $O(1)$ and consider separately the solution in the cell's interior and in a boundary layer around the separatrices. Note that the separatrices correspond to $\psi=0$ and that, away from the stagnation points, $\psi$ is a convenient coordinate with, for $\psi$ small, $\psi$ being proportional ot the distance from the separatrix. As in the homogenisation problem \citep[e.g.][]{chil79,rose-et-al}, the boundary-layer thickness scales like $\Pe^{-1/2}$; thus the cell interior is defined by $|\psi| \gg \Pe^{-1/2}$ while the boundary layer corresponds to $|\psi|=O(\Pe^{-1/2})$.

\subsection{Interior problem}

Introducing  the expansions
\beq \lab{expphif}
\phi = \phi_0 + \Pe^{-1/4} \phi_1 + \Pe^{-1/2} \phi_2 + \cdots
\inter{and}
f = f_0 + \Pe^{-1/4} f_1 + \Pe^{-1/2} f_2 + \cdots
\eeq
for the eigenfunction and eigenvalue into \eqn{eig1}, we obtain
\begin{eqnarray}
- \bu \cdot \nabla \phi_0 &=& 0, \lab{intphi0} \\
- \bu \cdot \nabla \phi_j + \bu \cdot \tbq \phi_{j-1} &=& 0, \ \ j=1,2,3, \lab{intphi1-3} \\
\nabla^2 \phi_0- \bu \cdot \nabla \phi_4 + \bu \cdot \tbq \phi_3 &=& f_0 \phi_0. \lab{intphi4}
\end{eqnarray}
The solution to \eqn{intphi0}--\eqn{intphi1-3} is straightforward: it corresponds to the expansion in powers of $\Pe^{-1/4}$ of $\exp(\Pe^{-1/4}\tbq \cdot \bx) \Phi(\psi)$, where $\Phi$ is arbitrary. Thus, in each quarter-cell,
\beq \lab{Phi0-3}
 \sum_{j=0}^3 \Pe^{-j/4} \phi_j= \e^{\Pe^{-1/4} \tbq \cdot \bx} \sum_{j=0}^3 \Pe^{-j/4} \Phi_j(\psi) + O(\Pe^{-1}),
 \eeq
where the functions $\Phi_j$ remain to be determined. In particular,
\beq \lab{Phi0}
\phi_0 = \Phi_0(\psi).
\eeq
Therefore, to leading order, the solution is constant along streamlines, in accordance with familiar averaging results \citep[cf.][]{rhin-youn,frei-went94}.
The higher-order terms in \eqn{Phi0-3} are not periodic, but periodicity is restored through the rapid variation of $\phi$ across the boundary layer. 

Eq.\ \eqn{intphi4} can be solved for $\phi_4$ provided that a solvability condition be satisfied. This solvability condition is obtained by integrating \eqn{intphi4} along a streamline. Noting that the third term on the left-hand side can be written as $
\bu \cdot \tbq \phi_3 = \bu \cdot \nabla (\cdots)$, where $\cdots$ denotes a polynomial of degree 4 in $\tbq \cdot \bx$ with $\psi$-dependent coefficients,
this condition is found to be
\beq \lab{int1}
\dt{}{\psi} \left( a(\psi) \dt{\Phi_0}{\psi} \right) = f_0 b(\psi) \Phi_0,
\eeq
where 
\beq \lab{ab}
a(\psi)=8 \left(E'(\psi)-\psi^2 K'(\psi) \right), \quad
b(\psi)=4 K'(\psi),
\eeq
and $K'$ and $E'$ are (complementary) complete elliptic integrals \citep[e.g.][]{dlmf}.  Details of the derivation of \eqn{int1}--\eqn{ab} are given in Appendix \ref{app:interior}. Note that \eqn{int1} can be recognised as an eigenvalue problem form of the diffusion equation obtained using averaging by \citet{rhin-youn}, \cite{frei-went94}, \citet{paul06} and others.

\begin{figure}
\begin{center}
\includegraphics[height=5cm]{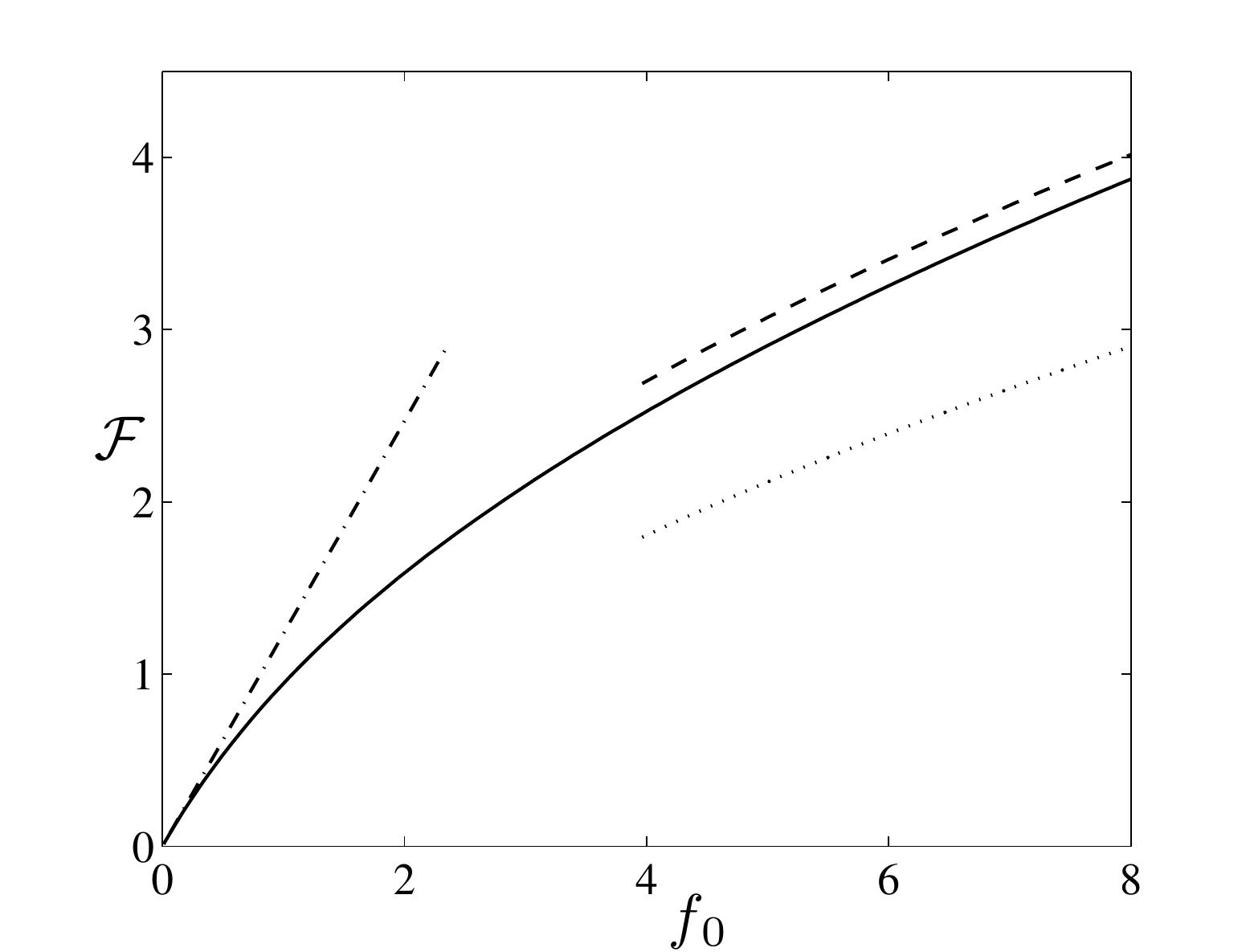} 
\caption{$\mathcal{F}$ as defined by \eqn{dir-neu} as a function of $f_0$. The numerical estimate of $\mathcal{F}$ (solid line) is compared with asymptotic approximations for $f_0 \ll 1$ (dash-dotted line) and for $f_0 \gg 1$ (dashed line and dotted line, corresponding to two- and one-term asymptotic approximations).}
\label{fig:Fcurly}
\end{center}
\end{figure}

As shown in Appendix \ref{app:interior}, the solution of \eqn{int1} that is well behaved at the centres $\psi=\mp 1$ of the cell  satisfies
\beq \lab{intbc2}
\dt{\Phi_0}{\psi}=\pm \frac{f_0}{2}  \Phi_0 \ \ \textrm{at} \ \ \psi= \mp 1.
\eeq
Eq.\ \eqn{int1} can be solved with this boundary condition and an arbitrary normalisation to find a linear relationship between $\Phi_0$ and its derivative near the separatrices: 
\beq \lab{dir-neu}
\dt{\Phi_0}{\psi} \sim \pm \mathcal{F}(f_0) \Phi_0 \ \ \textrm{as} \ \ \psi \to 0^\mp.
\eeq
This defines the function $\mathcal{F}(f_0)$  (Dirichlet-to-Neumann map) which in practice needs to be computed numerically. The boundary-layer analysis carried out in the next section determines the value of $\mathcal{F}(f_0)$ for a given $\bq$ and hence gives the leading-order approximation $f_0(\bq)$ to the eigenvalue.

The form of $\mathcal{F}(f_0)$ obtained by solving \eqn{int1}--\eqn{intbc2} numerically
is shown in Figure \ref{fig:Fcurly}. Note that the fact that $\mathcal{F}(f_0)$ is positive implies that the solution decays away from the separatrices towards the centres of the cells. 
The asymptotic behaviour of $\mathcal{F}$ for large and small $f_0$ is useful. Computations detailed in Appendix \ref{app:interior} give the following:
\begin{eqnarray}
\mathcal{F}(f_0) &\sim& \frac{\pi^2 f_0}{8} \quad \textrm{as} \ \ f_0 \to 0, \lab{f00} \\
\mathcal{F}(f_0) &\sim& \frac{\sqrt{2}\lambda}{4}  \left( 1 + \frac{\alpha}{\log \lambda}\right) \quad \textrm{as} \ \ f_0 \to \infty. \lab{f0infty}
\end{eqnarray}
In \eqn{f0infty}, $\lambda$ is a function of $f_0$ defined as the solution of 
\beq \lab{weber}
\lambda^2=4 f_0 \log \lambda,
\eeq
given explicitly in terms of  a Lambert function in \eqn{lambert}, and 
$\alpha$ is a constant given in \eqn{alpha}. The crude approximation
\beq \lab{f0inftycrude}
\mathcal{F}(f_0) \sim \frac{\left(f_0 \log f_0 \right)^{1/2}}{2}  \quad \textrm{as} \ \ f_0 \to \infty
\eeq
is readily derived from \eqn{f0infty} by neglecting the $O(\lambda/\log \lambda)$ term and using the leading-order approximation $\lambda \sim (2 f_0 \log f_0)^{1/2}$ to the solution of \eqn{weber}. This approximation is  very poor, however, with a relative error decreasing to $0$ only as $1/\log(\log f_0)$. The asymptotic approximations \eqn{f00}, \eqn{f0infty} and \eqn{f0inftycrude} are compared with the numerical solution in Figure \ref{fig:Fcurly}. This comparison validates the approximations and shows the importance of the logarithmic corrections to  \eqn{f0inftycrude}.

\subsection{Boundary layer and matching} \label{sec:Ibl}

In the boundary layer surrounding the separatrices, rescaled variables need to be introduced. Following \citet{chil79}, we let
\beq \lab{zeta}
\zeta = \mp \Pe^{1/2} \psi \inter{and} \sigma = \int_0^l |\nabla \psi| \, \d l, 
\eeq
where $l$ is the arclength along the separatrices.  The sign in \eqn{zeta} is chosen  such that $\zeta > 0$ in the interior of the quarter-cells. As detailed in Appendix \ref{app:bl1}, $0 < \sigma < 8$ parameterises the boundary of each quarter-cell, with $\sigma=0,\, 2, \, 4,\, 6$ at the corners (see Figure \ref{fig:cellpicture}).

The eigenvalue $f$ and eigenfunction $\phi$ are expanded as in \eqn{expphif}, with the latter now regarded as a function of $\zeta$ and $\sigma$. Introducing into \eqn{eig1} gives
\begin{eqnarray}
\partial^2_{\zeta\zeta}\phi_0 - \partial_\sigma \phi_0 &=& 0, \lab{bl1} \\
\partial^2_{\zeta\zeta}\phi_1 - \partial_\sigma{\phi_1} &=& - \frac{\bu \cdot \tilde{\bq}}{|\bu|^2} \phi_0  \lab{bl2} \\
\partial^2_{\zeta\zeta}\phi_2 - \partial_\sigma \phi_2 &=& - \frac{\bu \cdot \tilde{\bq}}{|\bu|^2} \phi_1  \lab{bl3}
\end{eqnarray}
Eq.\ \eqn{bl1} has the constant solution
\[
\phi_0 = \mathrm{const.} = \Phi_0(0),
\]
where $\Phi_0(0)$ is the limiting value of the leading-order interior solution on the separatrices. This indicates that the interior solution $\Phi_0(\psi)$ is the same in all quarter-cells. The problem posed by \eqn{bl2}--\eqn{bl3} is identical to the so-called Childress problem that arises in the computation of the effective diffusivity \citep{chil79}. It was solved in closed form by \citet{sowa87} using a Wiener--Hopf technique and is discussed further in Appendix \ref{app:bl1}. The key result is that
\beq \lab{dir-neu2}
\partial_\zeta \phi_1 \to 0 \quad  \textrm{and} \quad  \dpar{\phi_2}{\zeta} \sim - \frac{\pi^2 \nu}{4} |\tilde{\bq}|^2  \Phi_0(0) \ \ \textrm{as} \ \ \zeta \to \infty,
\eeq
where $\nu$ is as given in \eqn{nu}.

The leading-order approximation to the eigenvalue $f(\bq)$ is now obtained by matching the interior and boundary solutions. Comparing \eqn{dir-neu} and \eqn{dir-neu2} and taking the relation between $\zeta$ and $\psi$ \eqn{zeta} into account leads to
\[
\mathcal{F}(f_0)= \frac{\pi^2 \nu}{4} |\tilde{\bq}|^2,
\]
and hence
\beq \lab{f0cellPe1/4}
f(\bq) \sim \mathcal{F}^{-1} \left(\frac{\pi^2 \nu}{4} |\tilde{\bq}|^2\right),
\eeq
where $\mathcal{F}^{-1}$ denotes the inverse of $\mathcal{F}$. This is the desired approximation to the eigenvalue $f(\bq)$ in regime I, with $\Pe^{1/4}\bq=O(1)$ as $\Pe \to \infty$. It is completely explicit apart from the requirement for numerical solution of the ODE \eqn{int1} in order to determine $\mathcal{F}$. It indicates, in particular, that $f(\bq)$ depends only on $|\bq|$ in this regime, hence $g(\bxi)$ depends only on $|\bxi|$. Thus dispersion is isotropic not only in the diffusive regime but in the entire regime I. As shown below, the anisotropy of the dispersion appears  in regime II, for $\bq \gg \Pe^{-1/4}$. 

\begin{figure}
\begin{center}
\includegraphics[height=5cm]{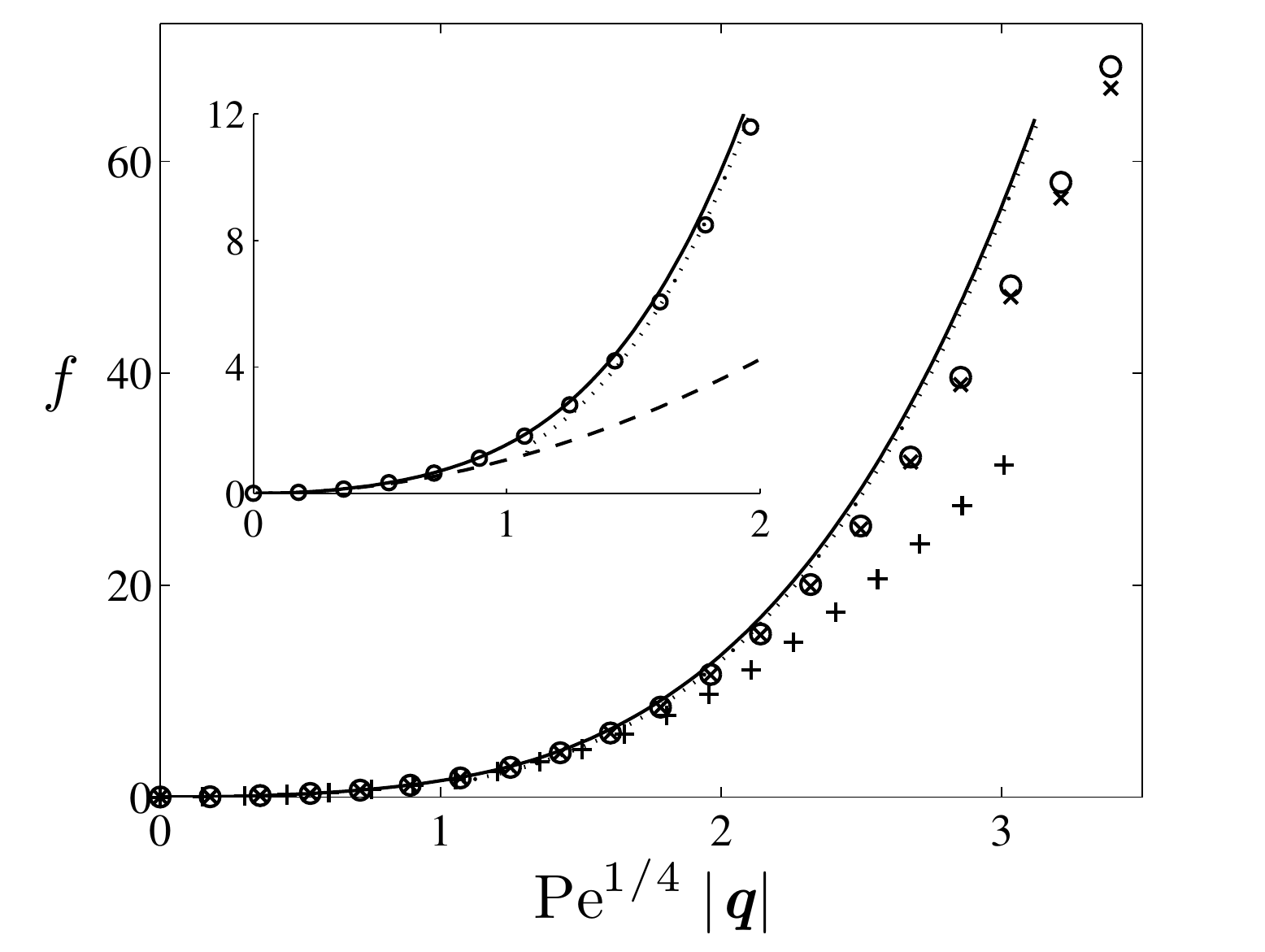} 
\caption{Eigenvalue $f(\bq)$ for the cellular flow   as a function of $|\bq|$ in  regime I, with $\Pe \gg 1$, $\Pe^{1/4} \bq = O(1)$. The asymptotic prediction \eqn{f0cellPe1/4} (solid line) is compared with numerical solutions of the eigenvalue problem for  $\Pe=5000$ with $\bq=|\bq|(1,1)/\sqrt{2}$ ($\circ$) and $\bq=(|\bq|,0)$  ($\times$), and for  $\Pe=500$ with $\bq=|\bq|(1,1)/\sqrt{2}$ (+). The approximation of \eqn{f0cellPe1/4} valid for $\Pe^{-1/4} \ll  \bq \ll 1$ that is deduced from \eqn{f0infty} is also shown (dotted curve). The diffusive approximation  \eqn{fdiff}, which holds for $\bq \ll \Pe^{-1/4}$, is shown in the inset magnifying the small-$|\bq|$ region (dashed line).} 
\label{fig:plotfqPe25}
\end{center}
\end{figure}

We have verified formula \eqn{f0cellPe1/4} by comparison with numerical estimates of $f(\bq)$ obtained by solving a discretisation of the eigenvalue problem \eqn{eig1} on a $1000^2$ grid (see Part I for details). The results are summarised in Figure \ref{fig:plotfqPe25}. The comparison between the numerical estimates obtained for different values of $\Pe$ (500 and 5000), and for different orientations of $\bq$ (parallel to $(1,1)$ and parallel to $(1,0)$) confirms the dependence of $f$ on $\Pe^{1/4} |\bq|$, the isotropy of the dispersion, and more generally  the validity of \eqn{f0cellPe1/4}.  

Formula \eqn{f0cellPe1/4} can be simplified further using the asymptotic approximations \eqn{f00} and \eqn{f0infty} of $\mathcal{F}$ for small and large argument. 
Using \eqn{f00}, \eqn{f0cellPe1/4} reduces to 
\beq \lab{fdiff}
f(\bq) \sim 2 \nu  |\tilde{\bq}|^2 = 2 \nu \Pe^{1/2} |\bq|^2 \quad \textrm{for} \ \ \bq \ll\Pe^{-1/4}
\eeq
This can be recognised as the diffusive approximation: the effective diffusivity deduced from \eqn{quadapp} recovers Soward's expression \eqn{effdiff}--\eqn{nu}. 
On the other hand, \eqn{f0inftycrude} gives
\beq \lab{fqlarge} 
f(\bq) \sim \frac{\pi^4 \nu^2  \Pe | \bq|^4}{16 \log \left(\Pe^{1/4} |\bq|\right)} \quad \textrm{for} \ \ \Pe^{-1/4} \ll \bq \ll 1.
\eeq
The latter approximation is  poor  because of the neglect of logarithmic terms, but it is useful in suggesting that $f(\bq)$ is proportional to $\Pe/\log \Pe$ when $\bq$ is not small. The two asymptotic approximations are shown in Figure \ref{fig:plotfqPe25}. For $\tilde{\bq} \gg 1$, we used a better approximation than \eqn{fqlarge} obtained by inverting \eqn{f0infty} numerically; this matches \eqn{f0cellPe1/4} accurately for $|\tilde{\bq}|  \gtrsim 1$.

\section{Regime II: $|\bq|=O(1)$}  \label{sec:II}
 
In this regime, the eigenfunction vanishes to leading order in the cell interiors. The problem is then entirely controlled by the behaviour inside the boundary layers around the separatrices. In contrast with the situation for $|\bq| = O(\Pe^{-1/4})$, the dynamics in the corners of the cells -- that is, near the stagnation points -- plays a crucial role. The eigenvalue $f(\bq)$ scales roughly like $\Pe$; it is therefore convenient to introduce
\beq \lab{tildef}
\f(\bq) = \Pe^{-1} f(\bq).
\eeq
Note that $\f(\bq)$ is not $O(1)$ but turns out to be $O(1 / \log \Pe)$; however, to obtain a reasonably accurate approximation to the eigenvalue, it is important to  capture logarithmic corrections: in what follows we therefore treat $1 / \log \Pe$ as an $O(1)$ quantity and neglect only terms that are algebraic in $\Pe^{-1}$. 

\subsection{Eigenvalue problem}

Away from the corners, the leading-order boundary-layer equation obtained from \eqn{eig1} using the variables \eqn{zeta} is
\beq \lab{blq1}
\partial_{\zeta\zeta}^2 \phi - \partial_\sigma \phi + \frac{\bu \cdot \bq}{|\bu|^2} \phi = \frac{\f}{|\bu|^2} \phi,
\eeq
where $\bu$ is evaluated on the separatrix. 
The solution can be written as
\beq \lab{varphi}
\phi = \e^{\bq \cdot \bx + \f H(\sigma)} \varphi,
\eeq
where
\beq \lab{H}
H(\sigma)=\frac{1}{2} \log \frac{2- \sigma}{\sigma} \ \ \textrm{for} \ \ 0 < \sigma < 2, \quad H(\sigma+2)=H(\sigma) 
\eeq
and the function $\varphi$ 
satisfies the heat equation
\beq \lab{heatphi}
\partial^2_{\zeta\zeta} \varphi - \partial_\sigma \varphi = 0 .
\eeq
See Appendix \ref{app:bl2} for details. Note that since $\phi$ is periodic, $\varphi$ is not, but there is a simple relation for the change in $\varphi$ under a translation that represents a map from one `$+$' (or one `$-$') cell to another. 

Eq.\ \eqn{blq1} breaks down near the corners $\sigma=0,\, 2,\, 4,\, 6$, where $\bu$ vanishes. There a different approximation to \eqn{eig1} needs to be considered; this provides a condition matching the form of $\phi$ downstream of the corners to its form upstream. The analysis of the corner region carried out in Appendix  \ref{app:bl2} gives this condition as
\beq \lab{cornerjump}
\lim_{\sigma \to k^+} \varphi(\sigma,\zeta) = (16 \Pe)^{-\f / 2} \zeta^{\f}  \lim_{\sigma \to k^-} \varphi(\sigma,\zeta) \ \ \textrm{for} \ k=0,\,2,\,4,\,6. 
\eeq

\begin{figure}
\begin{center}
\includegraphics[height=6cm]{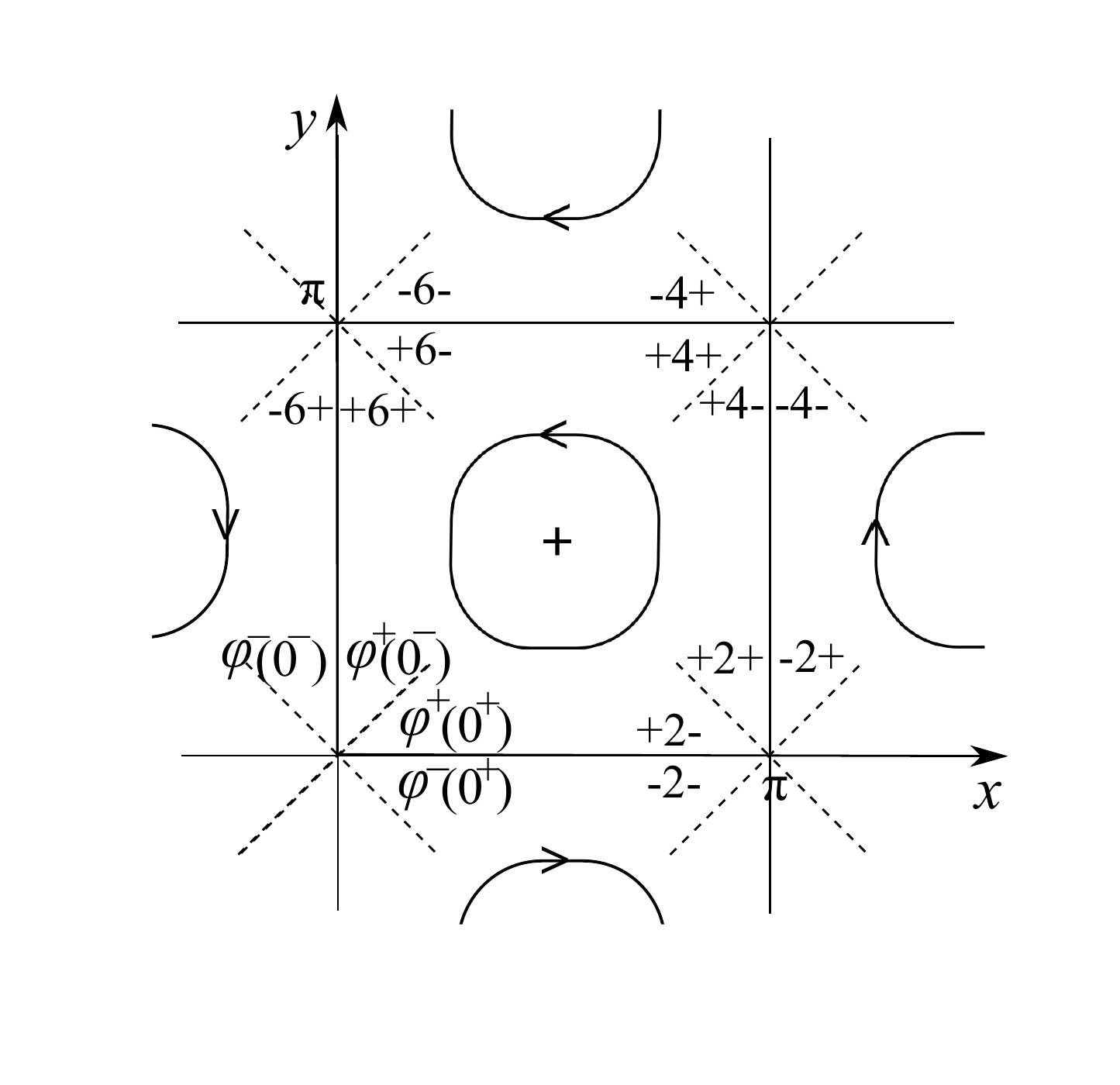} 
\vspace{-1cm}
\caption{Set-up of the boundary-layer analysis in regime II. The function $\varphi^+(\sigma,\zeta)$ and $\varphi^-(\sigma,\zeta)$ denote $\varphi$ respectively inside and outside of the separatrix around the central quarter-cell. The heat equation \eqn{heatphi} relates the functions $\varphi^\pm(\sigma,\zeta)$ immediately upstream of each corner, that is, for $\sigma=0^-,\, 2^-, \, 4^-,\, 6^-$, to 
the corresponding functions immediately downstream of each corner, that is, for $\sigma=6^+,\, 0^+, \, 2^+,\, 4^+$. These functions are indicated explicitly in the lower left corner $\sigma=0$ (omitting the dependence on $\zeta$), and symbolically in the other corners with, for instance, $+2-$ denoting $\varphi^+(2^-,\zeta)$.} 
\label{fig:qcell}
\end{center}
\end{figure}

Eqs.\ \eqn{heatphi}--\eqn{cornerjump}, together with with the jump conditions between $[0,2\pi]^2$ cells implied by the $2\pi$-periodicity of $\phi$, form an eigenvalue problem with $\varphi$ in each cell as eigenfunction and $\f$ as eigenvalue. It is in fact sufficient to consider  a single quarter-cell, say the $+$ cell centred centred at $(\pi/2,\pi/2)$: if $\varphi^+(\sigma,\zeta)$ and  $\varphi^-(\sigma,\zeta)$ denote $\varphi$ in the boundary layer inside and outside this cell (so that $\varphi^-$ straddles the four quarter-cells of type $-$ adjacent to the cell of type $+$, see Fig.~\ref{fig:qcell}), the periodicity of $\phi$ implies that the value of $\varphi$ in all other cells can be deduced from $\varphi^\pm$.  Furthermore, since solving \eqn{heatphi} between corners provides a map between $\varphi^\pm$ immediately downstream of each corner and $\varphi^\pm$ immediately upstream of the next corner, that is, between  $\varphi^\pm(k^+,\zeta)$ and $\varphi^\pm(k+2^-,\zeta)$, the problem can be formulated entirely in terms of $\varphi^\pm(k^+,\xi),\, k=0,\,2\,,4\,,6$. Defining a vector
\beq \lab{bvarphi}
\bvarphi(\zeta) = (\varphi^+(0^+,\zeta), \varphi^-(0^+,\zeta), \varphi^+(2^+,\zeta), \varphi^-(2^+,\zeta),\cdots,\varphi^-(6^+,\zeta))^\mathrm{T}
\eeq 
grouping these 8 functions, we show in Appendix \ref{app:bl2} that the problem can be written as
\beq \lab{matevalue}
(16 \Pe)^{\f/2} \bvarphi = \mathcal{L}(\bq,\f) \bvarphi.
\eeq
Here $\mathcal{L}(\bq,\f)$ is an $8 \times 8$ matrix, given explicitly in \eqn{matL}, whose entries are simple  linear integral operators. 

Let $\mu(\bq,\f)$ be the principal eigenvalue of $\mathcal{L}(\bq,\f)$:
\beq \lab{matLevalue}
\mathcal{L}(\bq,\f) \bvarphi = \mu(\bq,\f) \bvarphi.
\eeq
 Then $\f$ and hence $f$ are found as a functions of $\Pe$ by solving
\beq \lab{16Pe}
(16 \Pe)^{\f/2}  = \mu(\bq,\f) .
\eeq
This is the main result of this section. It gives the rate function $f$ as $\Pe$ times the solution $\f$ of the nonlinear equation \eqn{16Pe}. 
Since  $\f \to 0$ as $\Pe \to \infty$, it is asymptotically consistent to solve this equation approximately for small $\f$, which yields the leading-order approximation
\beq \lab{16Pesimp}
f(\bq) = \Pe \, \f(\bq) \sim \frac{2 \Pe}{\log \Pe} \log \mu(\bq,0).
\eeq
However, as mentioned earlier, this approximation is poor since it makes a relative error of order $O(1 / \log \Pe)$; it is therefore preferable to use \eqn{16Pe} instead. A heuristic improvement on \eqn{16Pesimp} retains the factor 16 inside the logarithm to read
\beq \lab{16Pesimp2}
f(\bq) = \Pe \, \f(\bq) \sim \frac{2 \Pe}{\log(16 \Pe)} \log \mu(\bq,0).
\eeq

\begin{figure}
\begin{center}
\includegraphics[height=5cm]{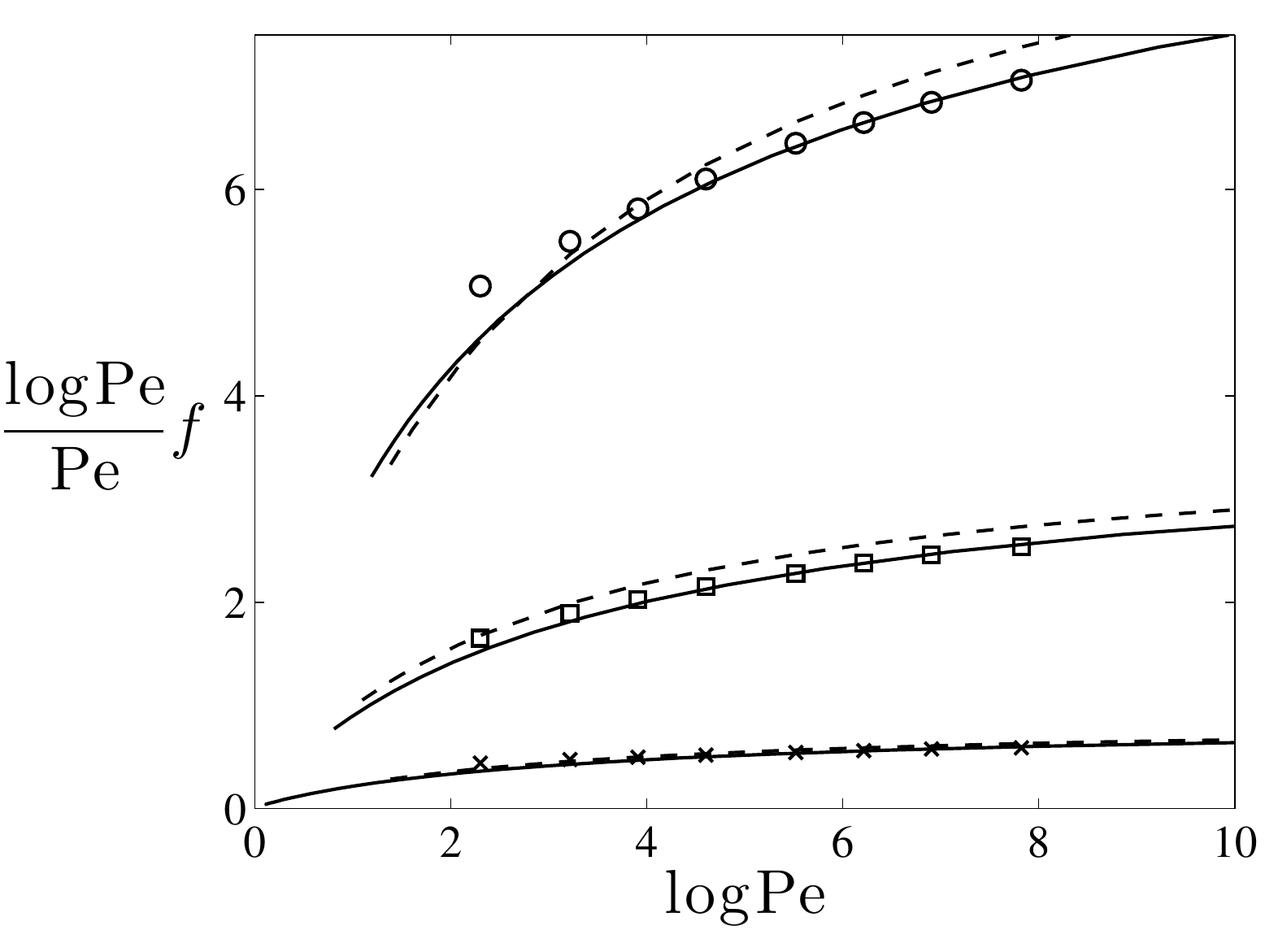} 
\caption{Scaled eigenvalue ${f(\bq)}$ as a function of $\Pe$ in regime II. The numerical solution of the eigenvalue problem (symbols) is compared with the asymptotic prediction \eqn{16Pe} (solid lines) for $q_1=q_2=0.5$ ($\times$), $1$ ($\Box$) and $2$ ($\circ$). As $\Pe \to \infty$, $\log \Pe f/ \Pe$ slowly tends to the limiting values $2 \log \mu(\bq,0)$ given here by $0.85$, $3.7$ and $10.0$. The heuristic formula \eqn{16Pesimp2} is indicated by the dashed lines.} 
\label{fig:logpeF}
\end{center}
\end{figure}

In practice, we can obtain $ \mu(\bq,\f)$ numerically by computing the eigenvalues of a discretised version of $\mathcal{L}(\bq,\f)$ for a range of $\f$, then deduce the corresponding values of $\Pe$ by solving \eqn{16Pe}. Alternatively, if $\f$ is to be estimated for a fixed $\Pe$, \eqn{16Pe} can be solved for $\f$ iteratively, starting with \eqn{16Pesimp}.
Figure \ref{fig:logpeF} demonstrates the accuracy of \eqn{16Pe} by comparing its prediction with the numerical solution of the full eigenvalue problem \eqn{eig1} for $f$ for values of $\Pe$ ranging from 10 to 2500 and for three different values of $q_1=q_2$. The figure indicates that \eqn{16Pe} is useful for values of $\Pe$ as small as 100. It also confirms the limited usefulness of the leading-order asymptotics \eqn{16Pesimp}: for $q_1=q_2=2$, for instance, the convergence of $\log \Pe f/ \Pe$ to its limiting value $2 \log \mu (\bq,0)=10.0$ is very slow so that exceedingly large $\Pe$ are required for an acceptable approximation. The heuristic formula \eqn{16Pesimp2}, although of the same formal accuracy, provides a clear improvement. 

%\begin{figure}
%\begin{center}
%\includegraphics[height=5cm]{fpe10002d} 
%\caption{Square root of the eigenvalue, $\sqrt{f(\bq)}$ for the cellular flow  for $\Pe =1000$ as a function of $\bq$. The numerical solution of the eigenvalue problem (solid contours and gray scale) is compared with the asymptotic prediction \eqn{16Pe} (dotted contours).} 
%\label{fig:fpe10002d}
%\end{center}
%\end{figure}

\begin{figure}
\begin{center}
\includegraphics[height=5cm]{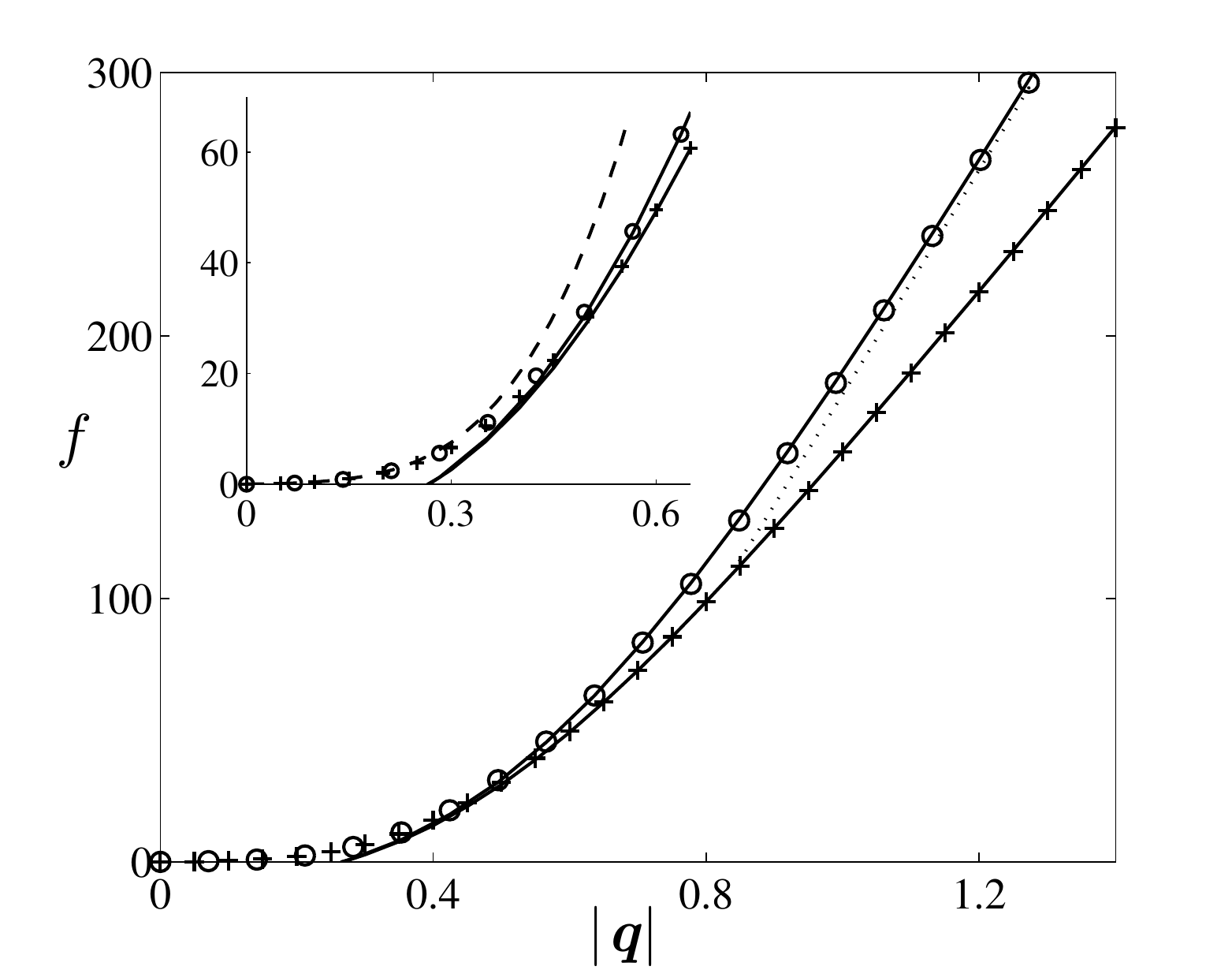} 
\caption{Eigenvalue $f(\bq)$ for the cellular flow  for $\Pe =1000$ as a function of $|\bq|$. The asymptotic prediction \eqn{16Pe} (solid line) is compared with numerical solutions of the full eigenvalue problem \eqn{eig1} with $\bq=|\bq|(1,1)/\sqrt{2}$ ($\circ$) and $\bq=(|\bq|,0)$  ($+$). The approximation  valid for $q_1, \, q_2 \gg 1$ that is deduced from \eqn{flargeq} is also shown (dotted line). The approximation \eqn{f0cellPe1/4} valid for $|\bq|=O(\Pe^{-1/4})$ is shown in the inset magnifying the small-$|\bq|$ region (dashed line).} 
\label{fig:fpe10001d}
\end{center}
\end{figure}

To illustrate the validity of \eqn{16Pe} over a broad range of $\bq$, we compare in Figure \ref{fig:fpe10001d} this prediction with numerical solutions along the lines $q_2=0$ and $q_1=q_2$ in the $\bq$-plane for $\Pe=1000$.
The  asymptotic prediction is virtually undistinguishable from the full numerical solution for $|\bq| \ge 0.5$. The figure confirms the anisotropy of dispersion in regime II. A two-dimensional plot of $f$ (see Figure 10 in Part I) indicates that the shape of constant-$f$ contours changes from circular for small values to straight segments given by $|q_1| + |q_2|=\textrm{const.}$ for large values. The behaviour is confirmed explicitly in the next subsection.

\subsection{Asymptotic limits}

The asymptotics of $f(\bq)$ for large and small $|\bq|$ is of interest. For simplicity, we consider the limit $|\bq| \ll 1$ in the approximation \eqn{16Pesimp}, that is, we neglect terms that are $O(1 / \log \Pe)$. 
A calculation detailed in Appendix \ref{app:rem2asy} relates the eigenvalue problem for $|\bq| \ll 1$ to that of the $|\bq|=O(\Pe^{-1/4})$ regime and yields
\beq \lab{fsmallq1}
f(\bq) \sim \frac{\pi^4 \nu^2 \Pe |\bq|^4}{4 \log \Pe},
\eeq
which matches the limiting form \eqn{fqlarge} of the $|\bq|=O(\Pe^{-1/4})$ regime. This shows that regimes I and II overlap in the region $\Pe^{-1/4} \ll |\bq| \ll 1$ where $f(\bq)$ is quartic in $|\bq|$.

For $|\bq| \gg 1$,  the eigenvalue problem \eqn{matevalue} can be greatly simplified by retaining only the dominant elements of the matrix $\mathcal{L}(\bq,\f)$. Specifically, assuming that both $|q_1|$ and $|q_2|$ are large, the eigenvalue $\mu(\bq,\f)$ can be approximated as
\beq \lab{muhat}
\mu(\bq,\f) \sim \e^{\pi(|q_1|+|q_2|)/2} \hat \mu(\f),
\eeq
where $\hat \mu(\f)$ is the eigenvalue of the ($\bq$-independent) scalar operator $\zeta^{\f} \mathcal{H}_-$, with $\mathcal{H}_-$  defined in \eqn{Hcurl}. This leads to the approximation
\beq \lab{flargeq}
f \sim \frac{\Pe}{\log(16 \Pe)} \left[\pi (|q_1|+|q_2|) + 2 \log \hat \mu(f/\Pe)\right].
\eeq
The first term in the square brackets is asymptotically dominant, but for practical values of $\Pe$ the second term needs to be taken into account (so that \eqn{flargeq} needs to be solved iteratively for $f$). Interestingly, \eqn{flargeq} can be related to the large-deviation statistics of random walks: a random walk on a two-dimensional lattice, with steps of size $\pm a$ taken with probability $1/2$ at time intervals $\tau$, is characterised by a large-deviation function 
\[
f(\bq)=\frac{1}{\tau} \log \left( \cosh(q_1 a) \cosh(q_2a) \right) \sim \frac{a}{\tau} (|q_1|+|q_2|) \quad \textrm{as} \ \ |\bq| \to \infty,
\]
assuming independent walks in the $x$- and $y$-directions.
Comparison with \eqn{flargeq} shows that for large $\Pe$ and large $|\bq|$, the dispersion by a cellular flow is equivalent to a random walk on the lattice of the hyperbolic stagnation points, with time intervals $\tau \propto \log \Pe/\Pe$ between the steps. This scaling is natural: $\Pe/\log \Pe$ is the time scale for both approaching the hyperbolic stagnation points along their stable manifold and for escaping from their neighbourhood along their unstable manifold (as consideration of the simple one-dimensional problems $\d X = \mp \Pe \sin X \, \d t + \sqrt{2} \, \d W$ readily confirms.) The physical interpretation is straightforward: since large values of $\bq$ correspond to large distances, $f(\bq)$ then describes the dispersion statistics of rare particles which travel anomalously fast away from their point of release. As Figure \ref{fig:trajplot} suggests, such particles move rapidly by remaining near the separatrices. The $O(\log \Pe/\Pe)$ time they take to pass through the regions surrounding  stagnation points is asymptotically larger than the $O(1/\Pe)$ time spent along the separatrix between these passages. From the current stagnation point a particle may move along the unstable manifold to one of the two connected stagnation points. As a result, the dynamics is approximated by that of a random walk with instantaneous jumps between neighbouring stagnation points. 

%These jumps are in practice alternately in the $x$ and $y$ directions (recall the geometry of separatrices shown in Figure \ref{fig:cellpicture}) but the large deviation statistics are the same as those for a random walk where jumps in the $x$ and $y$ directions are simply independent, rather than alternating. 

%Note that \eqn{flargeq} assumes both $|q_1|$ and $|q_2|$ large. For $|q_1| \gg 1$ but $|q_2|=O(1)$, corresponding to dispersion along the $x$-axis, it is also possible to simplify \eqn{16Pe}. This leads to an expression for $f(\bq)$ that is linear in $q_1$ but has a non-trivial dependence on $q_2$ that needs to be determined numerically. An analogous simplification naturally applies to the 
%complementary regime $|q_1|  =O(1)$ but $|q_2| \gg 1$. We omit the details of the asymptoticsolution in these two cases. 

Figure \ref{fig:fpe10001d} confirms the asymptotics \eqn{flargeq} by comparing its prediction with a numerical estimate for $f(\bq)$ for $\Pe=1000$ and $q_1=q_2$. The asymptotics is accurate for $q_1=q_2 \gtrsim 1$, not surprisingly perhaps since the large parameter is in fact $\min(\exp(\pi q_1),\exp(\pi q_2))$ (see \ref{app:rem2asy}). The asymptotics does not apply in the case $q_2=0$ also shown in the Figure and more generally if either $q_1$ or $q_2$ is $O(1)$; it is not difficult to obtain a simplified formula for this case, but this requires the eigenvalue of an operator more complicated than $\zeta^\f \mathcal{H}_-$. 
Figure \ref{fig:fpe10001d} also illustrates the switchover between regime I and regime II that  occurs in the range $\Pe^{-1/4} \ll |\bq| \ll 1$ where the  approximations \eqn{f0cellPe1/4} and \eqn{16Pe} are both valid and overlap.

We emphasise that the validity of the approximation given here for $|\bq| \gg 1$ is limited: for $|\bq|=O(\Pe)$ or larger, terms of \eqn{eig1} that are neglected in our boundary-layer treatment, most obviously the term $|\bq|^2 \phi$, become important. For $|\bq| \gg 1$ this term dominates so that $f(\bq) \sim |\bq|^2$, corresponding to a purely diffusive behaviour. Physically, this describes the statistics of particles that are so far from their release point that advection, with the limits imposed by the finite velocity, can no longer be the dominant mechanism of dispersion.  The transition between our regime II and this diffusion-dominated regime can be expected to take place in a third regime such that $|\bq|=O(\Pe)$ (up to logarithmic corrections); we leave the study of this regime for future work.

%[[JV: Peter, if we really really wanted to sort out this regime, Alexandra \& I have looked at it in the case $q_2=0$; I expect that what we did extends to $q_2 \not= 0$, though the case $q_1=q_2$ is special. Note that for large $\f$, the eigenvalue $\hat \mu$ of $\zeta^\f \mathcal{H}_-$ is given by $\exp((\f \log \f - \f)/2)$, with eigenfunction of the form $\exp(\f h(\f^{-1/2} \zeta))$ which peaks for some $\zeta=O(\f^{1/2})$ (the derivation is rather cute). ]] 

\section{Rate function}  \label{sec:rate}
  
In this section, we express the asymptotic results of \S\S\,\ref{sec:I}--\ref{sec:II} in terms of the rate function $g(\bxi)$. This is straightforward, since it only involves taking the Legendre transform of the (semi-)analytic formulas obtained for $f(\bq)$; it is useful however because, according to \eqn{largedevi}, the rate function directly provides the form of the scalar concentration. 
 
\begin{figure}
\begin{center}
\includegraphics[height=5cm]{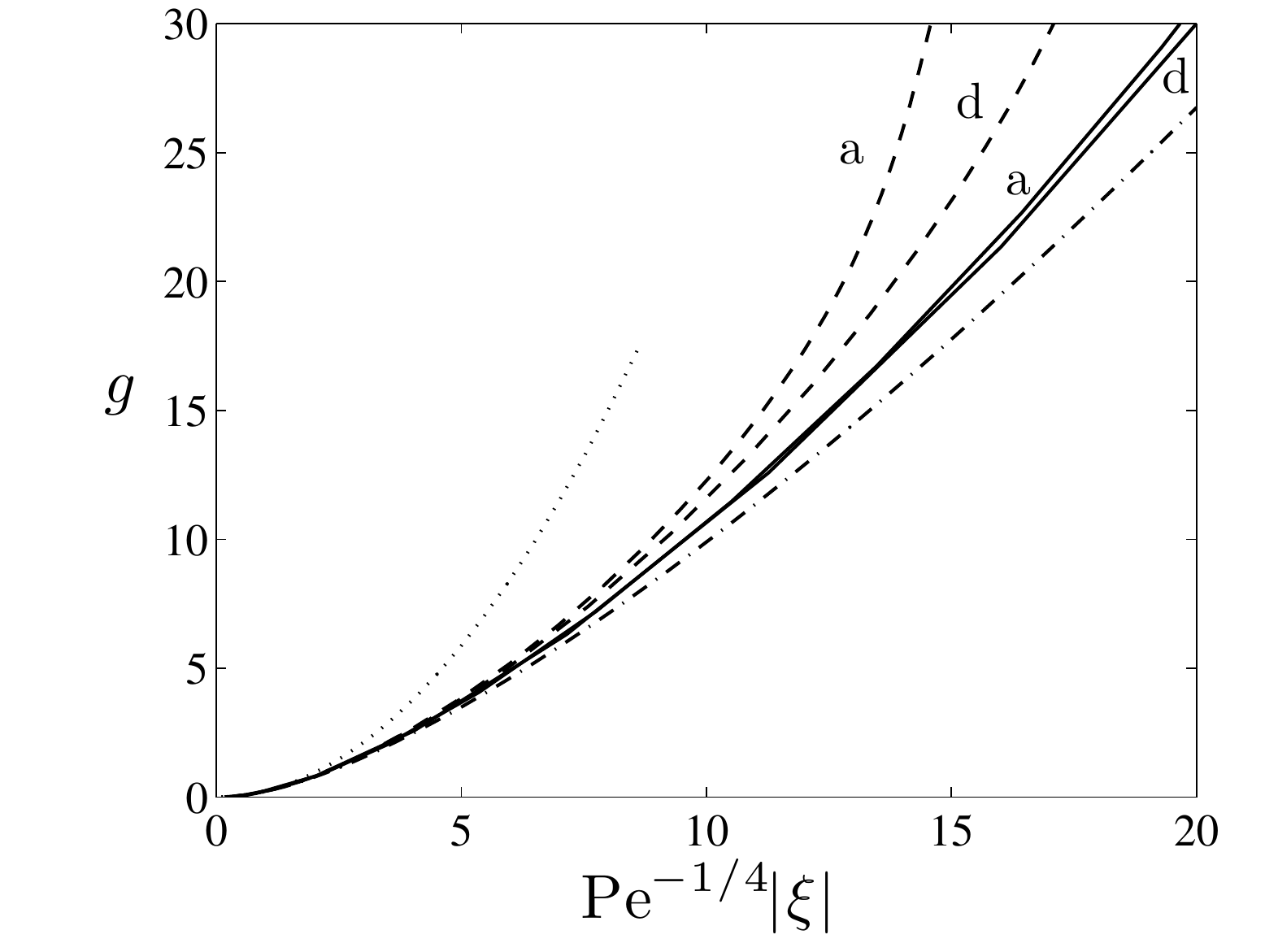} 
\caption{Rate function $g(\bxi)$ as a function of $\Pe^{-1/4} |\bxi|$ for $\Pe=100$ (dashed lines) and $\Pe=500$ (solid lines). Numerical estimates for $\bx=|\bxi|(1,1)/\sqrt{2}$ (labelled by `d' for diagonal) and for $\bxi=|\bxi|(1,0)$ (labelled by `a' for axis) are compared with the asymptotic approximation \eqn{gsmallxi} (dash-dotted line). The quadratic diffusive approximation \eqn{quadapp} is also shown (dotted line).} 
\label{fig:gsmallxi}
\end{center}
\end{figure}

The two regimes I and II identified for $f(\bq)$ naturally have counterparts for $g(\bxi)$. Regime I, which assumes $\bq=O(\Pe^{-1/4})$, is valid for $|\bq| \ll 1$ and yields $f(\bq)=O(1)$, is readily seen to correspond to $\bxi=O(\Pe^{1/4})$ and hold for $|\bxi| \ll \Pe/\log \Pe$. The Legendre transform of \eqn{f0cellPe1/4} corresponds to a rate function of the form
\beq \lab{gsmallxi}
g(\bxi) \sim \mathcal{G}(\Pe^{-1/4} | \bxi|),
\eeq
where the function $\mathcal{G}$, essentially the Legendre transform of $\mathcal{F}^{-1}$ in   \eqn{f0cellPe1/4} can be computed numerically. This prediction is verified in Figure \ref{fig:gsmallxi} which shows $g(\bxi)$ as a function of the scaled variable $\Pe^{-1/4} |\bxi|$ for $\Pe=100$ and $500$ and two different orientations of $\bxi$. As predicted, for $\Pe^{-1/4} |\bxi|$ not too large, the curves obtained by numerical solution of the eigenvalue problem collapse and match that obtained from the asymptotic solution. The diffusive approximation, with the quadratic $g$ given in \eqn{quadapp}, is also shown. The physical implications of the results for regime I derived from $f(\bq)$ can be reiterated based on the form of $g(\bxi)$: dispersion is isotropic in regime I, and the diffusive approximation considerably underestimates the dispersion, with $g(\bxi)$ much flatter than quadratic away from $|\bxi| \ll \Pe^{1/4}$, corresponding to exponentially higher concentrations than predicted by the effective diffusivity. The spatial distribution of the passive-scalar concentration, governed by the eigenfunction $\phi(\bx,\bxi)$, is shown in Part I: in regime I, the concentration has a non-trivial distribution in the cell interior with similar values in the boundary layer around the separatrices. As $\bxi$ increases from $0$, the interior concentration changes from near uniform to almost zero, with the boundary layer containing essentially all the scalar for $|\bxi| \gg \Pe^{1/4}$.

\begin{figure}
\begin{center}
\includegraphics[height=5cm]{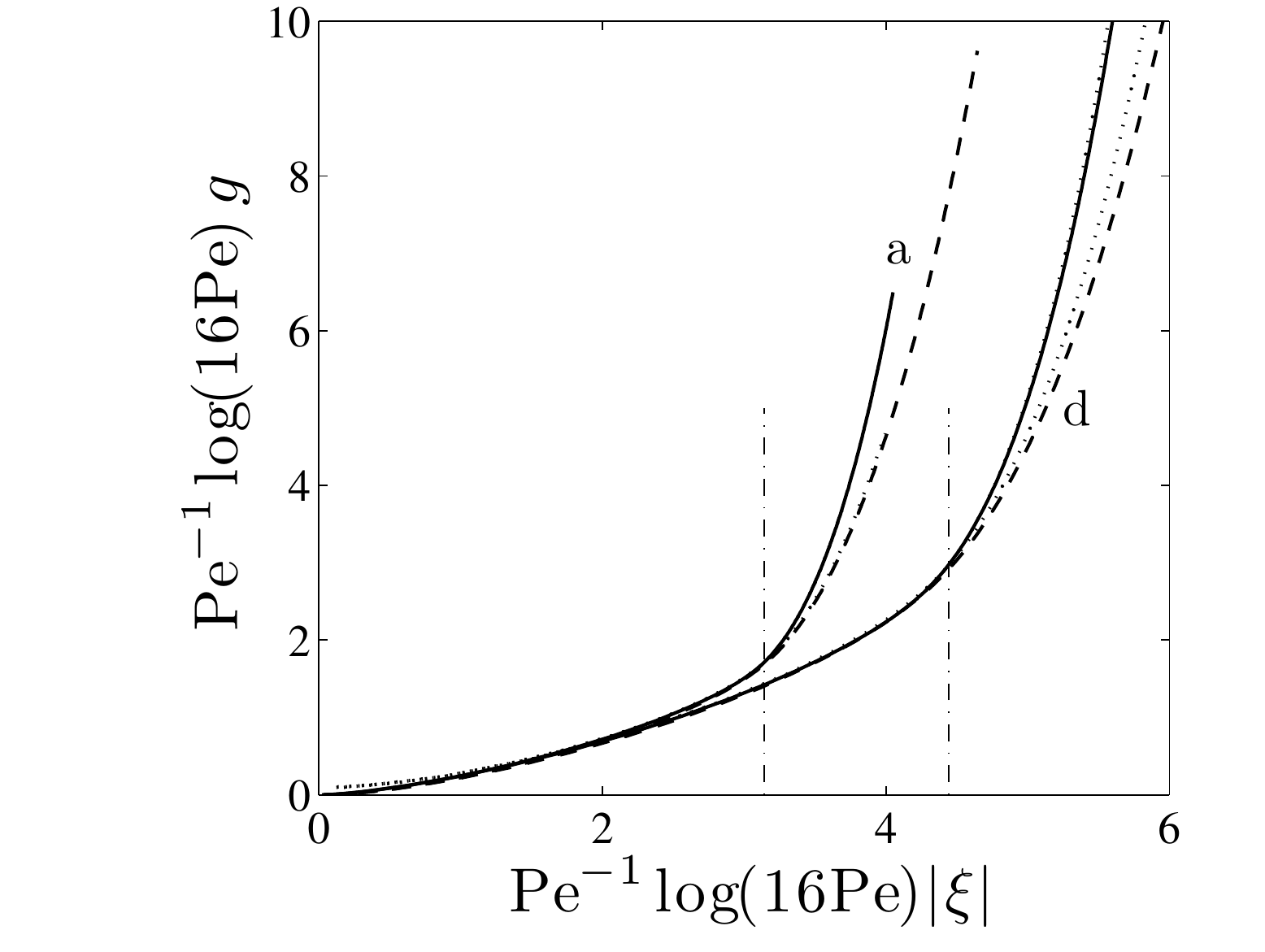} 
\caption{Rate function $g(\bxi)$ in the scaling of Regime II for for $\Pe=100$ (dashed lines) and $\Pe=500$ (solid lines). Numerical estimates for $\bx=|\bxi|(1,1)/\sqrt{2}$ (labelled by `d' for diagonal) and for $\bxi=|\bxi|(1,0)$ (labelled by `a' for axis) are compared with the asymptotic approximation derived from \eqn{16Pe} (dotted line, shown for $\bxi$ along the diagonal only). The dash-dotted vertical segments indicate the vertical asymptotes \eqn{verasy} predicted for $\Pe \to \infty$.} 
\label{fig:glargexi}
\end{center}
\end{figure}

The scaling of $g(\bxi)$ is regime II can be obtained from \eqn{16Pesimp2} with the caveat that neglect of some logarithmic terms limits the accuracy of the expression derived in this manner. The orders of magnitude $\bq=O(1)$ and $f(\bq)=O(\Pe/\log \Pe)$ associated with regime II correspond to  
$\bxi=O(\Pe/\log \Pe)$ and $g(\bxi)=O(\Pe/\log \Pe)$. More specifically, it follows from \eqn{16Pesimp2} that the rate function has the form
\beq \lab{glargexi}
 g(\bxi) \sim \frac{\Pe}{\log(16 \Pe)} \tilde\mathcal{G}\left(\frac{\log(16 \Pe)}{\Pe} \bxi\right)
\eeq
for some function $\tilde \mathcal{G}$ deduced from $\mu(\bq,0)$. As \eqn{16Pesimp2} itself this is an asymptotically inconsistent expression, taking into account the factor 16 inside the logarithms while neglecting other terms of the same, $O(1/\log \Pe)$ order. (The rate function deduced from $f$ obtained by solving \eqn{16Pe} gives a better approximation, but it is transcendental in  the P\'eclet number.)  Figure \ref{fig:glargexi} illustrates the behaviour of $g$ in regime II by showing the same numerical results as in Figure \ref{fig:gsmallxi} but with $g$ and $\bxi$ scaled according to \eqn{glargexi}. For $\Pe^{-1} \log(16 \Pe) \lesssim 1$, the pairs of curves corresponding to the same $\bxi$ but different P\'eclet numbers ($100$ and $500$) collapse, consistent with \eqn{glargexi}. For somewhat larger $\bxi$, the logarithmic corrections neglected matter, and the approximation derived from \eqn{16Pe} then provides the required approximation.

The most striking feature in the behaviour of $g(\bxi)$ is the abrupt increase once $|\bxi|$ exceeds a certain threshold. This implies that the concentration of a passive scalar drops suddenly for distances larger than $t$ times this threshold and, in practice, means that scalar is effectively localised to a finite support. This feature is captured by expression \eqn{flargeq} which provides an approximation of $f(\bq)$ for large $\bq$. Ignoring the correction term involving $\hat \mu(f/\Pe)$, the Legendre transform of \eqn{flargeq} implies that
\beq \lab{verasy}
g(\bxi) \to \infty \quad \textrm{as} \ \ \max\left( |\xi_1|, |\xi_2| \right) \to \xi_* = \frac{\pi\Pe}{\log(16 \Pe)}. 
\eeq
This provides an approximation for the size (and square shape) of this finite support for $\Pe \to \infty$. The finite support is entirely as expected for random walks with finite jumps separated by finite time intervals \citep[e.g.][]{kell04} which, as we have argued above, applies in the large $\Pe$ regime. For large-but-finite $\Pe$, the increase of $g (\bxi)$ with $|\bxi|$ is in fact smooth; it is encoded in $\hat \mu(f/\Pe)$ and for $|\xi_1|$ or $|\xi_2|$ substantially larger than $\xi_*$, by the form of $f(\bq)$ in the third regime $\bq=O(\Pe)$. However, the scalar concentrations corresponding such values of $\bxi$ (that is, such distances to the scalar-release point) are very small indeed.

\section{Conclusion} \label{sec:conc}

The large-deviation approach developed in Part I and extended here makes it possible to capture the tails in the distribution of a passive scalar released in the cellular flow \eqn{psi}.
It goes much further than the homogenisation approach classically applied to this problem: while homogenisation describes only the Gaussian core of the scalar distribution characterised by $|\bx|=O(t^{1/2})$, the large-deviation approach gives the prediction of a dependence in $\exp(-t g(\bx/t))$ valid for  much larger distances $|\bx| = O(t)$. It furthermore provides a method for determining the rate function $g$ by solving a family of eigenvalue problems. 

The asymptotic analysis of these eigenvalue problems in the large-P\'eclet-number limit reveals two district regimes in the dispersion. It is useful to summarise the predictions in these regimes using dimensional variables. Regime I holds for moderately large distances from the release point, specifically distances that satisfy $|\bx| \ll U t / \log \Pe$. It is characterised by an isotropic concentration that is broader than Gaussian and is controlled by the exchanges between the cell interiors across the separatrix regions. Regime II holds for  $ (\kappa^3 U/a^3)^{1/4} t \ll |\bx| = O(U t / \log \Pe) \ll U t$, is anisotropic and characterised by a sharp decrease of the concentration when $\max(|x|,|y|)$ approaches the specific value  $\pi U t / \log(16 \Pe)$. Regime II describes the statistics of particles that remain near the separatrix at all times. The main factor limiting the concentration is then  the passage through the stagnation points; near the sharp concentration decrease, the evolution is analogous to that of a random walk on the lattice formed by the stagnation points. For distances larger still, the concentration is exceedingly small and controlled by a different physical process in which molecular diffusion plays a key part; we do not analyse the corresponding regime III in this paper.

While this paper is entirely devoted to dispersion of passive scalars, the results are also relevant to problems involving reacting scalars. Specifically, the speed of propagation of fronts in models such as the FKPP equation turns out to be controlled by the large-deviation rate function $g(\bxi)$ for the corresponding passive-scalar problem \citep{gart-frei,frei85}.
%. Specifically, the speed $v_{\be}$ of propagation in direction $\be$ is the solution of $g(v_{\be} \be)= \alpha$  
Thus the asymptotic results of this paper can serve as a basis for new predictions for the propagation speed of fronts in cellular flows as studied, e.g., by \citet{abel-et-al02} and \citet{novi-ryzh}. These predictions are presented in  papers by \citet{tzel-v14b,tzel-v14a}; these also discuss regime III which turns out to be important for large reaction rates.

We conclude by noting that the form of the large-deviation theory used in this paper is an additive one, in the sense that it applies to SDEs with additive noise (and is therefore very close to Cram\'er's original theory for the sum of random numbers). Its multiplicative counterpart, exemplified by SDEs with multiplicative noise, is also relevant to the passive-scalar problem. It applies to the finite-time Lyapunov exponents which measure the rate of stretching experienced by line elements in a flow: for sufficiently mixing flows, their statistics obey a large-deviation principle and, remarkably, there is a significant part of parameter space in which the corresponding rate function controls the decay rate of the variance of passive scalars in such flows \citep[e.g.][]{tsan-et-al05a,hayn-v05}.

\smallskip

\noindent
\textbf{Acknowledgments.} The authors thank A. Tzella for useful discussions. JV acknowledges support from grant EP/I028072/1 from the UK Engineering and Physical Sciences Research Council. 

\appendix

\section{Derivation details for $|\bq|=O(\Pe^{-1/4})$}

\subsection{Interior solution} \label{app:interior}

We obtain the solvability condition \eqn{int1} by integrating \eqn{intphi4}  along streamlines. To do this, we introduce the time-like coordinate $s$ such that 
\beq \lab{char}
\dt{}{s}=\bu \cdot \nabla, \inter{i.e.} \d s =  \frac{\d x}{\sin x \cos y} = - \frac{\d y}{\cos x \sin y},
\eeq
that is used in conjunction with the value of the streamfunction $\psi$.
We first note that
\[
\oint \left( \bu \cdot \nabla \phi_4 + (\bu \cdot \tbq ) \phi_3 \right) \, \d s= 0,
\]
where the integration is along a streamline,
can be deduced from \eqn{Phi0-3}, specifically that $\phi_3$ is a sum of products of functions of $\tbq \cdot \bx$ and functions of $\psi$,  and that fact that $\oint \bu \cdot \nabla f(\bx) \, \d s =0$ for any function $f(\bx)$. Integrating \eqn{intphi4}  along a streamline then gives
\beq \lab{int0}
\oint \nabla^2 \phi_0 \, \d s = f_0 \oint  \phi_0 \, \d s .
\eeq
Now, following \citet{rhin-youn}, we use the arclength $\d l = |\nabla \psi| \d s$ to compute
\[
\oint \nabla^2 \phi_0 \, \d s = \dt{}{\psi} \int\!\!\int \nabla^2 \phi_0 \, \d x \d y = \dt{}{\psi} \oint \nabla \phi_0 \cdot \d \boldsymbol{l} = \dt{}{\psi} \left( \oint |\nabla \psi| \, \d l \, \dt{\phi_0}{\psi} \right)
\]
and reduce \eqn{int0} to the form \eqn{int1}, where
\beq
a(\psi) = \oint |\nabla \psi| \d l = \oint |\nabla \psi|^2 \, \d s \quad \textrm{and} \quad
b(\psi) = \oint \frac{\d l}{|\nabla \psi|} = \oint \d s
\eeq
can be recognised respectively as the circulation and the orbiting time around streamlines.

The explicit expressions \eqn{ab} for $a(\psi)$ and $b(\psi)$ are obtained as follows. To compute $b(\psi)$, we eliminate $y$ from \eqn{char} using the constancy of $\psi =-\sin x \sin y$ and compute
\[
b(\psi)= 2 \int_{\sin^{-1} \psi}^{\pi-\sin^{-1} \psi} \frac{\d x}{\sqrt{\sin^2 x - \psi^2}} = 4 K'(\psi),
\]
where $K'(\psi)=K(\sqrt{1-\psi^2})$ is a complete elliptic integral of the first kind \citep{dlmf}, and we have temporarily assumed that $0 \le \psi \le 1$. For $a(\psi)$, we observe that
\[
a(\psi) =  \int \!\! \int \nabla^2 \psi \, \d x \d y = -2  \int \!\! \int \psi \, \d \psi \d s = -2 \int \psi \d \psi \oint \d s = -2 \int \psi b(\psi) \, \d \psi,
\]
and hence that
\[
\dt{a}{\psi} = -2 \psi b(\psi).
\]
This equation can be integrated: using formula (19.4.2) in \citet{dlmf} we obtain
\[
a(\psi)=8 \left(E'(\psi)-\psi^2 K'(\psi) \right),
\]
where $E'(\psi)=E(\sqrt{1-\psi^2})$ is a complete elliptic integral of the second kind.

The following properties of $a(\psi)$ and $b(\psi)$ are useful:
\begin{eqnarray}
&a(0)=8, \qquad   b(\psi) \sim 4 \log(4/|\psi|)  \ \  \textrm{as} \ \psi \to 0, & \lab{ab0} \\
&a(\psi) \sim 4\pi(1\pm \psi) \  \textrm{as} \ \psi \to \mp 1, \qquad  b(\mp 1) =2\pi.& \lab{ab1}
\end{eqnarray}
In particular, using \eqn{ab1}, a Frobenius expansion shows that solutions of \eqn{int1} bounded at $\psi=\mp 1$ have the form
\[
\Phi_0 = C (1 + f_0(1 \pm \psi)/2 + O\left((1\pm\psi)^2)\right), 
\]
where $C$ in arbitrary constant
(the other solutions have a logarithmic singularity). This leads to the boundary condition \eqn{intbc2}.

We now consider the solution of \eqn{int1} in the limits $f_0 \to 0$ and $f_0 \to \infty$ and derive the asymptotic expressions \eqn{f00}--\eqn{f0infty} for $\mathcal{F}(f_0)$.   
For $f_0 \to 0$, $\Phi_0=1+O(f_0)$; introducing this in the right-hand side of \eqn{int1}, integrating  and imposing boundedness gives
\[
\dt{\Phi_0}{\psi} \sim \frac{f_0}{a(\psi)} \int^\psi_{\pm 1} b(\psi') \, \d \psi'
\]
and hence 
\[
\mathcal{F}(f_0) \sim \frac{f_0}{a(0)} \int_0^1 b(\psi) \, \d \psi = \frac{\pi^2 f_0}{8}.
\]
For $f_0 \to \infty$, it is convenient to introduce
\beq \lab{r}
r(\psi)=\frac{a(\psi)}{\Phi_0(\psi)}\dt{\Phi_0}{\psi}
\eeq
which satisfies the Riccati equation
\beq \lab{riccati}
\dt{r}{\psi}=f_0 b(\psi) - \frac{r^2}{a(\psi)}.
\eeq
The solutions of interest decay with $|\psi|$ and are approximated by
\[
r = \pm \left(f_0 a(\psi) b(\psi)\right)^{1/2}
\]
away from $\psi=0$. (A boundary layer of $O(f_0^{-1})$ width appears around the centres $\psi=\mp 1$ so that the boundary condition  \eqn{intbc2} can be satisfied.)

Near $\psi=0$, \eqn{riccati} is approximated as
\beq \lab{r1}
\dt{r}{\psi}=4 f_0 \log(4/|\psi|) - \frac{1}{8} r^2 + O(f_0 \psi^2).
\eeq
using \eqn{ab0}.
A dominant-balance argument suggests to introduce
\beq \lab{PsiR}
\Psi = \lambda \psi \inter{and} R = r/\lambda,
\eeq
where $\lambda$ satisfies 
\beq
\lambda^2 = 4 f_0 \log \lambda
\eeq
Note that this equation has the closed-form solution
\beq \lab{lambert}
\lambda = \exp \left( -\frac{W_\mathrm{m}(-1/(2f_0))}{2}\right)
\eeq
in terms of the Lambert function $W_\mathrm{m}$ \citep[e.g.][]{dlmf}, and the approximate solution
\[
\lambda \sim \left(2 f_0 \log f_0 \right)^{1/2} \quad \textrm{as} \ \ f_0 \to \infty.
\]
 Introducing \eqn{PsiR} transforms \eqn{r1} into
\beq \lab{R}
\dt{R}{\Psi}=  1 - \frac{R^2}{8} + \frac{\log(4/|\Psi|)}{\log\lambda} + O(1/(\lambda \log \lambda)),
\eeq
where we assume $\Psi=O(1)$. We solve this equation perturbatively: the expansion
\beq \lab{R2}
R = \pm  2\sqrt{2} + \frac{R_1}{\log \lambda} + O\left(1/(\lambda \log \lambda) \right),
\eeq
where $R_1(\psi)$ remains to be determined, 
satisfies \eqn{R} to leading order. At the next order, we find
\[
\dt{R_1}{\Psi} = \pm \frac{\sqrt{2}}{2} R_1 + \log \frac{4}{|\Psi|}.
\]
The solution that is bounded as $|\Psi| \to \infty$ takes the form
\[
R_1 = \mp \sqrt{2} \log (|\Psi|/4) \mp \sqrt{2} \e^{\sqrt{2}|\Psi|/2} \, \mathrm{Ei}(\sqrt{2}|\Psi|/2),
\]
where $\mathrm{Ei}$ is the exponential integral \citep[e.g.][]{dlmf}. Evaluating at $\Psi=0$ gives 
\[
R_1(0)=\pm \frac{\sqrt{2}}{2} \left(3 \log 2 + 2 \gamma \right) ,
\]
where $\gamma$ is Euler's constant. After introducing this result into \eqn{R2}, we find from \eqn{PsiR}, \eqn{r} and \eqn{ab0} that
\beq \lab{alpha}
\mathcal{F}(f_0) \sim \frac{\sqrt{2}\lambda}{4}  \left( 1 + \frac{\alpha}{\log \lambda}\right), \inter{where} \alpha = \frac{3 \log 2 + 2 \gamma}{4} = 0.8084682178\cdots.
\eeq

\subsection{Boundary-layer solution} \label{app:bl1}

The boundary layer equations \eqn{bl2}--\eqn{bl3} are essentially identical to those to be solved to compute the effective diffusivity using a homogenisation approach. Thus the solution follows closely \citet{chil79} and \citet{sowa87} and is detailed here for completeness \citep[see also][]{chil-sowa89}.

Since the solution is identical in the quarter-cells with the same sense of flow rotation, we concentrate on the $+$ quarter-cell $[0,\pi] \times [0,\pi]$ and on the $-$ cell $[0 ,\pi] \times [\pi, 2 \pi]$. We note that
\begin{eqnarray*}
0 <  \sigma = 1  - \cos x < 2 \ \ \textrm{for} \ \ y=0, &&
2 < \sigma = 3 -\cos y < 4   \ \textrm{for} \ \ x=\pi, \\
4 <  \sigma = 5  + \cos x < 6 \ \ \textrm{for} \ \ y=\pi, &&
6 < \sigma = 7 +\cos y < 8   \ \textrm{for} \ \ x=0.
\end{eqnarray*}
in the + quarter-cell, while
\begin{eqnarray*}
0 <  \sigma = 1  - \cos x < 2 \ \ \textrm{for} \ \ y=2 \pi, &&
2 < \sigma = 3 -\cos y < 4   \ \textrm{for} \ \ x=\pi, \\
4 <  \sigma = 5  + \cos x < 6 \ \ \textrm{for} \ \ y=\pi, &&
6 < \sigma = 7 +\cos y < 8   \ \textrm{for} \ \ x=0.
\end{eqnarray*}
in the $-$ quarter-cell.
Eqs.\ \eqn{bl2}--\eqn{bl3} are solved in these two quarter-cells. This leads to solutions $\phi_k^\pm(\sigma,\zeta),\, k=1,2$, that need to be matched across the separatrice $\zeta=0$. Using periodicity, the matching conditions are found to be
\begin{eqnarray*}
\phi^+_k(\sigma,0)=\phi^-_k(\sigma,0) \ \ \textrm{and} \ \ \partial_\zeta \phi^+_k(\sigma,0)=- \partial_\zeta \phi^-_k (\sigma,0) &&   \textrm{for} \ \ 0 < \sigma < 2, \ 4 < \sigma < 6, \\
\phi^+_k(\sigma,0)=\phi^-_k(\sigma+4,0) \ \ \textrm{and} \ \ \partial_\zeta \phi^+_k(\sigma,0)=- \partial_\zeta \phi^-_k (\sigma+4,0) &&   \textrm{for} \ \ 2 < \sigma < 4, \    6 < \sigma < 8, 
\end{eqnarray*}
with the $\phi_k^\pm$ periodic with period 8 (see Fig.~\ref{fig:cellpicture}). 

Using that $\phi_0=\Phi_0(0)$ is a constant, \eqn{bl2} is written explicitly as
\beq \lab{phi1s}
\partial^2_{\zeta\zeta} \phi^\pm_1 - \partial_\sigma \phi_1^\pm = - \tilde q_1 F(\sigma) \mp \tilde q_2 F(\sigma-2),
\eeq
where 
\beq \lab{F}
F(\sigma) = \left\{ \begin{array}{ll}
(2 \sigma - \sigma^2)^{-1/2} &  \textrm{for} \ \ 0 < \sigma < 2 \\
0 & \textrm{for} \ \ 2 < \sigma < 4
%\\
%- (2 (\sigma-4) - (\sigma-4)^2)^{-1/2} &  \textrm{for} \ \ 4 < \sigma < 6 \\
%0 & \textrm{for} \ \ 6 < \sigma < 8 
\end{array} \right. \quad \textrm{and} \quad F(\sigma+4)=-F(\sigma).
\eeq
For convenience, we have used the linearity of  \eqn{bl2}--\eqn{bl3} to set $\phi_0=\Phi_0(0)=1$ temporarily.

It follows from \eqn{phi1s}--\eqn{F} that $\phi_1^\pm (\sigma+4,\zeta)=-\phi_1^\pm(\sigma,\zeta)$ and so the matching conditions become
\begin{eqnarray}
\phi^+_1(\sigma,0)=\phi^-_1(\sigma,0) \ \textrm{and}  \ \partial_\zeta \phi^+_1(\sigma,0)=- \partial_\zeta \phi^-_1 (\sigma,0) &&   \textrm{for}  \ 0 < \sigma < 2, \ 4 < \sigma < 6, \lab{bcpm1}\\
\phi^+_1(\sigma,0)=-\phi^-_1(\sigma,0)  \ \textrm{and} \ \partial_\zeta \phi^+_1(\sigma,0)=\partial_\zeta \phi^-_1 (\sigma,0) &&   \textrm{for} \ 2 < \sigma < 4, \ 6 < \sigma < 8, \lab{bcpm2}
\end{eqnarray}

Defining $G(\sigma)$ by 
\beq \lab{G}
G'(\sigma) = F(\sigma) \inter{and} \int_0^8 G(\sigma) \, \d \sigma = 0,
\eeq
that is,
\[
G(\sigma)=\left\{
\begin{array}{ll}
\sin^{-1}(\sigma-1) &  \textrm{for} \ \ 0 < \sigma < 2, \\
\pi/2   & \textrm{for} \ \ 2 < \sigma < 4
\end{array}
\right., \quad G(\sigma+4)=-G(\sigma),
\]
the solution to \eqn{phi1s}--\eqn{bcpm2} can be written as
\beq \lab{phivarrho}
\phi_1^\pm(\sigma,\zeta) =  \tilde q_1 \left(G(\sigma) + \varrho(\sigma,\zeta)\right) \pm \tilde q_2 \left(G(\sigma-2) + \varrho(\sigma-2,\zeta)\right), 
\eeq
where $\varrho(\sigma,\zeta)$ satisfies
\begin{eqnarray*}
\partial^2_{\zeta\zeta} \varrho - \partial_\sigma \varrho = 0, && \partial_\zeta \varrho(\sigma,0)=0 \ \ \textrm{for} \ \ 0 < \sigma < 2, \   4 < \sigma < 6, \\
&& \varrho(\sigma,0)=-G(\sigma) \ \ \textrm{for} \ \ 2 < \sigma < 4, \ 6 < \sigma < 8.
\end{eqnarray*}
Since  $G(\sigma)=\pi/2$ for $ 2 < \sigma < 4$ and $G(\sigma)=-\pi/2$ for $6 < \sigma < 8$, 
\beq \lab{theta}
\varrho(\sigma,\zeta)=-\frac{\pi}{2}\theta(\sigma+2,\zeta), 
\eeq
where $\theta(\sigma,\zeta)$ satisfies
\begin{eqnarray*}
\partial^2_{\zeta\zeta} \theta - \partial_\sigma \theta = 0, 
&& \theta(\sigma,0)=-1 \ \ \textrm{for} \ \ 0 < \sigma < 2, \\
&& \theta(\sigma,0) = 1 \ \ \textrm{for} \ \   4 < \sigma < 6, \\
&& \partial_\zeta \theta(\sigma,0)=0 \ \ \textrm{for} \ \ 2 < \sigma < 4, \ \  6 < \sigma < 8, 
\end{eqnarray*}
with $\theta \to 0$ as $\zeta \to \infty$. This is the problem solved in closed form by \citet{sowa87} using a Wiener--Hopf technique. 

The solution \eqn{phivarrho} can now be introduced into Eq.\ \eqn{bl3} satisfied by $\phi_2$ in order to obtain $\partial_\zeta \phi_2$ as $\zeta\to \infty$. We first rewrite \eqn{bl3} as 
\beq \lab{aaa}
\partial^2_{\zeta\zeta} \phi_2^\pm - \partial_\sigma \phi_2^\pm = - \left[\tilde q_1 F(\sigma) \pm \tilde q_2 F(\sigma-2)\right] \phi_1^\pm,
\eeq
and note that the leading-order behaviour of the solution as $\zeta \to \infty$ is controlled by the average in $\sigma$:
\[
\phi^\pm_2 \sim \bar{\phi}^\pm_2 = \frac{1}{8} \int_0^8 \phi^\pm_2 \, \d \sigma \ \ \textrm{as} \ \  \zeta \to \infty.
\]
Introducing \eqn{phivarrho} into the average of \eqn{aaa} and using \eqn{G} and symmetries reduces this equation to
\[
\partial^2_{\zeta\zeta} \bar \phi_2^\pm = - \frac{|\tilde{\bq}|^2}{4} \int_0^2 F(\sigma) \varrho(\sigma,\zeta) \, \d \sigma.
\] 
Integrating with respect to $\zeta$ then gives
\[
\partial_\zeta \bar \phi^\pm_2 = - \frac{|\tilde{\bq}|^2}{4} \int_0^2 F(\sigma) \int_0^\infty \varrho(\sigma,\zeta) \, \d \zeta \d \sigma = -\frac{\pi |\tilde{\bq}|^2}{4} \int_0^\infty \varrho(0,\zeta) \, \d \zeta,
\]
on using that $\partial_\sigma \int_0^\infty \varrho(\sigma,\zeta) \, \d \zeta = - \partial_\zeta \varrho(\sigma,0) = 0$ for $0< \sigma < 2$. We finally obtain \eqn{dir-neu2} using \eqn{theta} above and formula (A.9) in \citet{sowa87}, namely
\[
\int_0^\infty \theta(2,\zeta) \, \d \zeta = - 2 \nu,
\]
where $\nu$ is defined by \eqn{nu}.

\section{Derivation details for $|\bq|=O(1)$} \label{app:reg2det}

\subsection{Boundary-layer solution} \label{app:bl2}

To solve \eqn{blq1}, we  note that $\partial_\sigma (\cdot) = \bu \cdot \nabla(\cdot) / |\bu|^2$, hence $\partial_\sigma \bx = \bu/|\bu|^2$, and that $H(\sigma)$ defined in \eqn{H} satisfies
\[
H(\sigma)=-\int_1^\sigma F^{2}(\sigma') \, \d \sigma' = -\int_1^\sigma \frac{\d \sigma'}{|\bu(\sigma')|^2}.
\]
The undifferentiated terms in \eqn{blq1} can therefore be integrated explicitly to obtain \eqn{varphi}--\eqn{heatphi}. The periodicity of $\phi$ and the symmetry of the system imply the relationships
\beq \lab{varphiper}
\varphi(x+ k \pi,y+l \pi) = \e^{-\pi (k q_1 + l q_2)} \varphi(x,y),
\eeq
for all integers $k, \, l$ with $k+l$ even. This makes it possible to deduce $\varphi$ in all the boundary layers from its form $\varphi^\pm$ on the interior and exterior sides of the boundary layer of the  quarter-cell with centre at $(\pi/2,\pi/2)$. 

We now obtain condition \eqn{cornerjump} governing the jump in $\varphi$ at each corner. For definiteness, let us consider the corner at $(0,0)$ of the $+$ quarter-cell with centre at $(\pi/2,\pi/2)$. Since $-x y \sim \psi = O(\Pe^{-1/2})$ near this corner, suitable rescaled coordinates are $\bX=\Pe^{1/4} \bx$; it terms of these, \eqn{eig1} reduces to
\beq \lab{corner}
X \partial_X \phi - Y \partial_Y \phi + \f \phi = 0
\eeq
to leading order in $\Pe$. The solution is 
\beq \lab{cornersoln}
\phi = X^{-\f} \Phi(XY)
\eeq
for some function $\Phi$ that is found by matching with the solution \eqn{varphi} valid away from the corner. Upstream of the corner, this matching is made in the limit $Y \to \infty$ with $XY=\zeta$ fixed; noting that $Y = \Pe^{1/4} y \sim \Pe^{1/4} (-2 \sigma)^{1/2}$ and $H(\sigma) \sim \log(-\sigma/2)/2$, we find 
\beq \lab{upstream}
\Phi(\zeta) =  (16 \Pe)^{-\f/4} \zeta^\f \lim_{\sigma\to 0^-} \varphi(\sigma,\zeta).
\eeq
Note that we retain the factor $\zeta^\f$ although it is asymptotically small since $\f \to 0$ as $\Pe \to \infty$. This is because this leads to logarithmic corrections to $f(\bq)$ which are not negligible for large-but-finite $\Pe$.

Downstream of the corner the analogous matching corresponds to the limit $X \sim \Pe^{1/4} (2 \sigma)^{1/2} \to \infty$ with $XY=\zeta$ fixed and leads to
\beq \lab{downstream}
\Phi(\zeta) =  (16 \Pe)^{\f/4}  \lim_{\sigma\to 0^+} \varphi(\sigma,\zeta)
\eeq
using that $H(\sigma) \sim \log(2/\sigma)/2$.
Comparing \eqn{downstream} with \eqn{upstream} leads to the jump condition \eqn{cornerjump} at $\sigma=0$. The same condition applies to all corners. 

We are now in position to write down the eigenvalue problem determining $\f$. The heat equation \eqn{heatphi} makes it possible to relate the values of $\varphi$ upstream of each corner to that downstream of the preceding corner. Specifically, integrating the heat equation for $0 < \sigma < 2$ gives
\begin{eqnarray} 
\varphi^+(2^-,\zeta) &=& \mathcal{H}_+ \varphi^+(0^+,\zeta) + \mathcal{H}_- \varphi^-(0^+,\zeta), \lab{phi+phi-} \\ 
\varphi^-(2^-,\zeta) &=& \mathcal{H}_- \varphi^+(0^+,\zeta) + \mathcal{H}_+ \varphi^-(0^+,\zeta), \lab{phi-phi+}
\end{eqnarray}
where $\mathcal{H}_\pm$ are the linear operators giving the `time'-$2$ flow of the heat equation and are defined by 
\beq \lab{Hcurl}
\left(\mathcal{H}_\pm h\right)(\zeta) = \frac{1}{\sqrt{8 \pi}} \int_0^\infty \e^{-(\zeta\mp \zeta')^2/8} h(\zeta') \, \d \zeta'
\eeq
for any function $h(\zeta)$.

Relations analogous to \eqn{phi+phi-} can be written down for $\varphi^\pm(4^-,\zeta)$, $\varphi^\pm(6^-,\zeta)$ and $\varphi^\pm(8^-,\zeta)=\varphi^\pm(0^-,\zeta)$. The jump condition \eqn{cornerjump} can then be used to eliminate $\varphi^\pm(2k^-)$ in favour of $\varphi^\pm(2k^+)$. For $\varphi^+$, this is straightforward: \eqn{cornerjump} gives
\[
\varphi^+(0^+,\zeta) = (16 \Pe)^{-\f/2} \zeta^\f \varphi^+(0^-,\zeta)
\]
and similar relations at the other 3 corners. For $\varphi^-$, this is somewhat more complicated: because $\varphi^-$ is defined in 4 different quarter-cells, the upstream profiles are not immediately available. The periodicity conditions\eqn{varphiper} can however be used to express 
them in terms of $\varphi^-(2k^-)$, see Figure \ref{fig:qcell}. This leads to the jump conditions
\begin{eqnarray*}
\varphi^-(0^+,\zeta) = (16 \Pe)^{-\f/2} \e^{\pi(q_1+q_2)} \zeta^\f \varphi^-(4^-,\zeta), && 
\varphi^-(2^+,\zeta) = (16 \Pe)^{-\f/2} \e^{\pi(q_2-q_1)} \zeta^\f \varphi^-(6^-,\zeta), \\
\varphi^-(4^+,\zeta) = (16 \Pe)^{-\f/2} \e^{-\pi(q_1+q_2)} \zeta^\f \varphi^-(0^-,\zeta), && 
\varphi^-(6^+,\zeta) = (16 \Pe)^{-\f/2} \e^{\pi(q_2-q_1)} \zeta^\f \varphi^-(2^-,\zeta).
\end{eqnarray*}
Gathering these results, the eigenvalue problem can be written in the vector form \eqn{matevalue} where the linear operator $\mathcal{L}$ is
\beq \lab{matL}
\mathcal{L}(\bq,\f) = \zeta^\f
\left( \begin{array}{cccccccc}
0 & 0 & 0 & 0 & 0 & 0 & \mathcal{H}_+ & \mathcal{H}_- \\
0 & 0 & ab \mathcal{H}_- & ab \mathcal{H}_+ & 0 & 0 & 0 & 0 \\
\mathcal{H}_+ & \mathcal{H}_- & 0 & 0 & 0 & 0 & 0 & 0 \\
0 & 0 & 0 & 0 & a^{-1}b \mathcal{H}_- & a^{-1}b \mathcal{H}_+ & 0 & 0 \\
0 & 0 & \mathcal{H}_+ & \mathcal{H}_- & 0 & 0 & 0 & 0 \\
0 & 0 & 0 & 0 & 0 & 0 & (ab)^{-1} \mathcal{H}_- & (ab)^{-1}  \mathcal{H}_+ \\
0 & 0 & 0 & 0 & \mathcal{H}_+ & \mathcal{H}_- & 0 & 0 \\
a b^{-1} \mathcal{H}_- & a b^{-1} \mathcal{H}_+ & 0 & 0 & 0 & 0 & 0 & 0 
\end{array}
\right),
\eeq
with $a=\e^{\pi q_1}$ and $b=\e^{\pi q_2}$.

\subsection{Asymptotic limits} \label{app:rem2asy}

In the limit $|\bq \ll 1$, the eigenvalue problem \eqn{matLevalue} or, equivalently, \eqn{blq1} can be solved by perturbation expansion in powers of $|\bq|$. Since $\f$ decreases rapidly with $|\bq|$ (like $|\bq|^4$ as is verified below), the right-hand side of \eqn{blq1} and the jumps \eqn{cornerjump} are negligible. Expanding $\phi$ in powers of $|\bq|$ then leads to the same sequence of equations \eqn{bl1}--\eqn{bl3} as considered in the $|\bq|=O(\Pe^{-1/4})$ regime. The solution is the same: to leading-order $\phi$ is a constant, and the slope of the solution as $\zeta \to \infty$ is related to this constant according to \eqn{dir-neu2}. This implies that
\beq \lab{dir-neu3}
\dpar{\bvarphi}{\zeta} \sim - \frac{\pi^2 \nu}{4} |\bq|^2 \bvarphi \ \ \textrm{as} \ \ \zeta \to \infty.
\eeq
This perturbative solution breaks down for large $\zeta$, however, since the constant leading-order $\phi$ is inconsistent with the decay requirement for $\bvarphi$. For large $\zeta$, the eigenfuction $\varphi$ takes an exponential form: 
\[
\bvarphi \propto \exp(-\lambda \zeta) \ \ \textrm{as} \ \ \zeta \to \infty
\] 
for some $\lambda$. Comparing with \eqn{dir-neu3} gives
\beq \lab{lambda}
\lambda = \frac{\pi^2 \nu}{4} |\bq|^2.
\eeq
It can be verified that the decay rate $\lambda$ is related to the eigenvalue $\mu(\bq,0)$ by
\beq \lab{lambdamu}
\mu(\bq,0)=\e^{2 \lambda^2}.
\eeq
Combining \eqn{16Pesimp}, \eqn{lambda} and \eqn{lambdamu} yields the approximation \eqn{fsmallq1}.

For $|q_1|, \, |q_2| \gg 1$, the eigenvalue problem \eqn{matLevalue} simplifies dramatically. For definiteness, we assume $q_1, \, q_2 > 0$. In this case, $a^{-1} b^{-1} \ll 1 \ll ab$ in \eqn{matL}. This implies that $\varphi_2$, the second component of $\bvarphi$, is its largest component, and that $\varphi_4 \ll \varphi_2 \ll \varphi_3$. Taking this into account reduces \eqn{matLevalue} to
\[
\mu \left(\begin{array}{c} \varphi_2 \\
\varphi_3 \end{array} \right)= \zeta^\f \left(\begin{array}{cc}
0 & ab \mathcal{H}_- \\
\mathcal{H}_- & 0 \end{array} \right) \left(\begin{array}{c} \varphi_2 \\
\varphi_3 \end{array} \right),
\]
and hence to
\[
\mu^2 \varphi_2 = ab \zeta^{2\f} \mathcal{H}_-^2 \varphi_2.
\]
Therefore, $\mu(\bq,\f) = (ab)^{1/2} \hat{\mu}(\f)$, where $\hat{\mu}(\f)$ is the eigenvalue of $\zeta^\f \mathcal{H}_-$ and \eqn{muhat} follows.

\bibliographystyle{agsm}

\end{document}